\documentclass[1p,times,preprint]{elsarticle}
\usepackage[english]{babel}
\usepackage{tensor}
\usepackage{graphicx}
\usepackage{amsmath}
\usepackage{amssymb}
\usepackage{amsfonts}
\usepackage{dcolumn}
\usepackage{bm}
\usepackage{xcolor}
\usepackage{ulem}
\usepackage{tikz}
\usepackage{subcaption}
\usepackage{comment}
\usepackage{verbatim}
\usepackage{fancyvrb}
\usepackage{cancel}
\usepackage{multirow}
\usepackage{lscape}
\usepackage{txfonts}
\usepackage{mathtools}
\usepackage{soul}
\usepackage{url}
\usepackage{longtable}
\usepackage{makecell}
\usepackage[pdftex]{pict2e}
\usepackage{microtype}
\newcolumntype{d}[1]{D{.}{.}{#1}}
\usepackage{color, colortbl} %%LK
\definecolor{Blue}{rgb}{0.8,1,1} %%LK
\usepackage[first=0,last=9]{lcg} %%LK
\def\bra<#1|{\mathinner{\langle\,{#1}\,\vert}} 
\def\ket|#1>{\mathinner{\vert\,{#1}\,\rangle}} 
\def\red|#1|{\mathinner{\!\vert\,{#1}\,\vert\!}}
\def\braket<#1>{\mathinner{\langle\,{#1}\,\rangle}} 
\def\redmem#1#2#3{  \left\langle #1 \left\Vert  
                  #2 \right\Vert #3 \right\rangle   }

\begin{document}
%%%%%%%%%%%%%%%%%%%%%%%%%%%%%%%%%%%%%%%%%%%%%%%%%%%%%%%%%%%%%%%%%%%%%%%%%%%%%%%

\title{ SECOND-ORDER RAYLEIGH-SCHR\"ODINGER PERTURBATION THEORY FOR THE GRASP2018 PACKAGE:
THREE-PARTICLE FEYNMAN DIAGRAM CONTRIBUTION TO CORE-VALENCE CORRELATIONS}
\date{\today}

\author[TFAI]{G. Gaigalas}
\address[TFAI]{Institute of Theoretical Physics and Astronomy, Faculty of Physics,
               Vilnius University, Saul\.{e}tekio Ave. 3, LT-10257 Vilnius, Lithuania}
\ead{gediminas.gaigalas@tfai.vu.lt}

\author[TFAI]{P. Rynkun}
\ead{pavel.rynkun@tfai.vu.lt}

\author[TFAI]{L. Kitovien\.{e}}
\ead{laima.radziute@tfai.vu.lt}

%%%%%%%%%%%%%%%%%%%%%%%%%%%%%%%%%%%%%%%%%%%%%%%%%%%%%%%%%%%%%%%%%%%%%%%%%%%%%%%
%
%   ABSTRACT
%
%%%%%%%%%%%%%%%%%%%%%%%%%%%%%%%%%%%%%%%%%%%%%%%%%%%%%%%%%%%%%%%%%%%%%%%%%%%%%%%
\begin{abstract}
In order to ascertain the precise atomic characteristics, 
it is important to incorporate a wide range of electron correlations 
in the computational analysis. However, it is important to note 
that undertaking such studies will result 
in substantial increases of the CSF expansions. 
The present publication, 
as well as the series of articles 
that have been published by G. Gaigalas, P. Rynkun, and L. Kitovienė, 
are dedicated to the analysis 
and resolution of the problem in question. 
In this series, a method was developed based 
on second-order perturbation theory to identify 
the most important core-valence, core, core-core, 
and valence-valence correlations. The method under 
discussion is based on a combination of the relativistic 
configuration interaction method and the stationary 
second-order Rayleigh-Schrödinger many-body perturbation 
theory in an irreducible tensorial form. In this study, 
the method is further expanded to encompass additional 
core-valence electron correlations of the third and fourth types. 
Conversely, the correlations that are not encompassed by 
perturbation theory are addressed in a conventional manner. 
This method can be used to calculate the properties 
of an atom or ion with any number of valence and core electrons. 
As an example of its application, the atomic calculations 
of the energy structure and lifetime for Ne~II are presented. 
\end{abstract}

%\pacs{31.15.A-, 31.15.ag, 31.15.aj, 31.15.am, 31.15.V-, 31.30.J-}
\begin{keyword}
configuration interaction \sep spin-angular integration \sep perturbation theory \sep tensorial algebra \sep valence-valence  correlations \sep core-valence correlations \sep core correlations \sep core-core correlations

%\PACS 31.15.-p \sep 31.15.Ne \sep 31.30.Jv \sep 03.65.Pm 

\end{keyword}
\maketitle

%\newpage

\section{Introduction}

In this work, a further development of the second-order Rayleigh-Schr\"odinger many-body perturbation theory (RSMBPT)~\cite{LindgrenBook:82} in an irreducible tensorial form~\cite{Gaigetal:2024CV,Gaigetal:2024C,Gaigetal:2024CC,Gaigetal:2025VV,Gaigetal:2025VVT} is presented for a computational package {\sc Grasp}~\cite{grasp2018,grasp2025} based on the multiconfiguration Dirac-Hartree-Fock (MCDHF) approximation and the relativistic configuration interaction (RCI) method~\cite{grantBook:07,Fisetal:16a,Jonetal:23a}. This combination greatly facilitates the selection of the configuration state function (CSF) space in a way that does not sacrifice computational accuracy but minimises the space itself in RCI method.

In this paper, the methodology is extended to core-valence correlations, which are conditioned by a three-particle Feynman diagram. Although in this case a large number of CSFs that represent these correlations do not enter the computation, for completeness of the methodology and for developing a general algorithm, we consider this extension to be important in the generation of the CSF space~\cite{Jonetal:23b}. Moreover this extension facilitates the use of the methodology for complex systems.

Similarly, as in our previous papers~\cite{Gaigetal:2024CV,Gaigetal:2024C,Gaigetal:2024CC,Gaigetal:2025VV,Gaigetal:2025VVT}, we have provided in this paper (Section 2) the analytical expression of a three-particle Feynman diagram describing the core-valence correlations with explanations. Since the spin-angular part of this diagram is much more complex, a significant part of the paper is devoted to calculating the spin-angular coefficients~\cite{Gaigalas_1997,Gaigalas:2026a} and the extension of library \texttt{librang}~\cite{Gaigalas:2022} (Section 3). The final expressions are presented in Section 4 of the paper. Section 5 contains calculations showing that the expressions and the examination of the core-valence correlations presented in the paper are correct and usable. Conclusions are presented in Section 6.

\section{Relativistic second-order effective Hamiltonian of an atom or ion in irreducible tensorial form for the remaining core-valence correlations}
\label{sec:seconOrder}

New Feynman diagrams are used to describe core-valence (CV) correlations, which were not available for core (C)~\cite{Gaigetal:2024C},
 core-core (CC)~\cite{Gaigetal:2024CC}, and valence-valence (VV)~\cite{Gaigetal:2025VV} correlations. We discuss this in detail below.

\subsection{The third type of core-valence correlations}
\label{sec:PT_Mano_Third}

The first two types of core-valence correlations are already described in the paper~\cite{Gaigetal:2024CV}.
The third type of core-valence correlations is presented through the Feynman diagram CV$_7$ from Fig.~\ref{CV_7} where all lines with the double arrow of diagram are renamed in the following way: $p' \equiv m$ and $m'=n=n'=p \equiv n$.
\begin{equation}
\label{eq:CVT-a} 
    (n_{a} \ell_{a})\, j_{a}^{2j_a+1} \, (n_{m} \ell_{m})\, j_{m}^{w_m} \, (n_{n} \ell_{n})\, j_{n}^{w_n} 
% \nonumber \\
   \rightarrow (n_{a} \ell_{a})\, j_{a}^{2j_a} \; (n_{m} \ell_{m})\, j_{m}^{w_m-1} \, (n_{n} \ell_{n})\, j_{n}^{w_n+2}.
\end{equation}

\begin{figure*}
\begin{center}
\setlength{\unitlength}{1mm}
%\begin{picture}(160,43)
\begin{picture}(180,43)(4,0)
\thicklines
\put(10,30){\line(0,1){10}}
\put(10,38){\vector(0,1){2}}
\put(10,40){\vector(0,1){2}}
\put(7.5,38){\makebox(0,0)[t]{\small{$m$}}}
\put(11.9,26.1){\makebox(0,0)[t]{\small{$n$}}}
\multiput(10,30)(1,0){10}{\circle*{0.35}}
%---------------------------------------
\put(10,20){\line(0,10){10}}
\put(22.5,24.5){\makebox(0,0)[r]{\small{$a$}}}
%\put(20,38){\vector(0,1){2}}
%\put(20,40){\vector(0,1){2}}
\put(12.5,22.5){\vector(1,1){2.4}}
\put(11.5,21.5){\vector(1,1){2}}
\put(20,26){\vector(0,-1){3}}
%vidurinijo vertikalioji linija
%\put(20,10){\line(0,10){30}}
\put(20,30){\line(-1,-1){8.7}}
\put(20,20){\line(0,10){10}}
\put(20,20){\line(-1,1){8}}
%\put(20,10){\vector(0,1){3}}
%\put(20,08){\vector(0,1){3}}
\put(14.5,25.5){\vector(-1,1){2}}
\put(13.5,26.5){\vector(-1,1){2.4}}
\put(12.5,20.4){\makebox(0,0){\small{$n^{\prime}$}}}
\multiput(20,20)(1,0){10}{\circle*{0.35}}
%---------------------------------------
\put(10,10){\line(0,1){10}}
\put(10,10){\vector(0,1){3}}
\put(7.5,13){\makebox(0,0){\small{$m^{\prime}$}}}
\put(10,08){\vector(0,1){3}}
\put(30,10){\line(0,1){20}}
\put(30,30){\vector(0,1){3}}
\put(30,28){\vector(0,1){3}}
\put(27.5,28){\makebox(0,0){\small{$p$}}}
\put(30,10){\vector(0,1){3}}
\put(30,08){\vector(0,1){3}}
\put(27.5,13){\makebox(0,0){\small{$p^{\prime}$}}}
\put(20,04){\makebox(0,0){$\text{CV}_{7}$}}
\put(37,37){\makebox(0,0) [l] {$\displaystyle{ = -
%\frac{1}{\overline{E}\left(K\right)-\overline{E}\left(K^{'}\right)}
	\sum_{k, k^{\prime}, x}~ \sqrt{\frac{\left[ x \right]}{\left[ k, k^{\prime} \right]}} ~\sum_{m, m^{\prime}}~\sum_{n, n^{\prime}}~\sum_{p, p^{\prime}}
	}$}}
\put(40,25){\makebox(0,0) [l] {$\displaystyle{ \times
		\left[\left[\left[\;  a^{\left( j_m \right) }  \times 
  \tilde a^{\left( j_{m^{\prime}} \right) } \; \right] ^{\left( k \right)} \times 
	\left[\; \tilde a^{\left( j_{n^{\prime}} \right) }  \times 
  a^{\left( j_{n} \right) } \; \right] ^{\left( x \right)} \right]^{\left( k^{\prime} \right)}	
  \times
	\left[\; a^{\left( j_{p} \right) }  \times 
  \tilde a^{\left( j_{p^{\prime}} \right) } \; \right] ^{\left( k^{\prime} \right)} \right]^{\left( 0 \right)}
	}$}}
\put(40,13){\makebox(0,0) [l] {$\displaystyle{ \times
\sum_{a}~\frac{1}
{\left( \varepsilon_{a}+\varepsilon_{p'}-\varepsilon_n-\varepsilon_p \right)}
	\left\{
    \begin{array}{ccc}
      j_{n} & j_{n'} & x \\
      k      & k'    & j_{a}  
     \end{array}
	\right\}  
	X_{k}(m a, m^{\prime} n^{\prime}) ~ X_{k'}(n p, a p^{\prime})}$}}
\end{picture}
\caption{The CV Feynman diagram of the second-order effective Hamiltonian for the third and fourth types of core-valence correlations 
$(n_{a} \ell_{a})\, j_{a}^{2j_a+1} \, (n_{m} \ell_{m})\, j_{m}^{w_m} \, (n_{n} \ell_{n})\, j_{n}^{w_n} 
   \rightarrow (n_{a} \ell_{a})\, j_{a}^{2j_a} \; (n_{m} \ell_{m})\, j_{m}^{w_m-1} \, (n_{n} \ell_{n})\, j_{n}^{w_n+2}$
and
$(n_{a} \ell_{a})\, j_{a}^{2j_a+1} \, (n_{m} \ell_{m})\, j_{m}^{w_m} \, (n_{n} \ell_{n})\, j_{n}^{w_n} \, (n_{p} \ell_{p})\, j_{p}^{w_p} 
   \rightarrow (n_{a} \ell_{a})\, j_{a}^{2j_a} \; (n_{m} \ell_{m})\, j_{m}^{w_m-1} \, (n_{n} \ell_{n})\,j_{n}^{w_n+1}  \; (n_{p} \ell_{p})\, j_{p}^{w_p+1}$.
}
\label{CV_7}
\end{center}
\end{figure*}

The three-particle Feynman diagram's CV$_7$ expression, like the diagrams described in the previous papers~\cite{Gaigetal:2024CV,Gaigetal:2024C,Gaigetal:2024CC,Gaigetal:2025VV,Gaigetal:2025VVT}, have an energy denominator
$D = \sum \left( \varepsilon_{\text{down}} - \varepsilon_{\text{up}} \right)$,
where $\varepsilon_{\text{down}}$ ($\varepsilon_{\text{up}}$) is the single-particle eigenvalue associated with the down- (up-)
orbital lines to (from) the lowest interaction line of the diagram. For example, the denominators for the CV$_7$ diagram are
\begin{equation}
\label{eq:denominator}
D = \left( \varepsilon_{a}+\varepsilon_{p'}-\varepsilon_n-\varepsilon_p \right) ,
\end{equation}
where index $a$ belong to $F$ set, and $p'$, $n$, $p$ belong to $F'$ set of orbitals~\cite{Gaigetal:2024CV}.

Also, the following notations are used in the expressions of this diagram (see Fig.~\ref{CV_7}):
\begin{equation}
\label{eq:deffX}
   X_{k}(i j, i' j') 
   = \redmem{\ell_i j_{i}}{\, C^{(k)} \,}{ \ell_{i'} j_{i'}}
     \redmem{\ell_j j_{j}}{\, C^{(k)} \,}{ \ell_{j'} j_{j'}} 
%\nonumber \\
%   \times \, 
R^{k}(n_i j_i \, n_jj_j, \, n_{i'}j_{i'} \, n_{j'}j_{j'} ) ,
\end{equation}
where $R^{k}\left(n_i j_i \, n_jj_j, \, n_{i'}j_{i'} \, n_{j'}j_{j'} \right)$ is the radial integral 
of electrostatic interaction between electrons~\cite[(89) and (90)]{Fisetal:16a} and 
$\redmem{\ell_i j_{i}}{\, C^{(k)} \,}{ \ell_{i'} j_{i'}}$ is the reduced matrix element of the irreducible tensor operator $C^{(k)}$ in $jj-$coupling.

This diagram has six operators of second quantization (three pairs of creation and annihilation operators) as was in the paper~\cite{Gaigetal:2025VVT}. This complicates finding the values of the spin-angular coefficient of this diagram and therefore makes the original program library \texttt{librang}~\cite{Gaigalas:2022} from {\sc Grasp} unusable in the most general case. But for this case the problem was already solved in the paper~\cite{Gaigetal:2025VVT} and the program library \texttt{librang}~\cite{Gaigalas:2022} was appropriately extended.

\subsection{The fourth type of core-valence correlations}
\label{sec:PT_Mano_Fourth}

\begin{figure*}
\begin{center}
\setlength{\unitlength}{1mm}
%\begin{picture}(160,43)
\begin{picture}(180,93)(8,0)
\thicklines
% First Feynman diagram y + 40
\put(10,80){\line(0,1){10}}
\put(10,88){\vector(0,1){2}}
\put(10,90){\vector(0,1){2}}
\put(7.5,88){\makebox(0,0)[t]{\small{$m$}}}
\put(11.9,76.1){\makebox(0,0)[t]{\small{$p$}}}
\put(15,82){\makebox(0,0){\small{$k$}}}
\multiput(10,80)(1,0){10}{\circle*{0.35}}
%---------------------------------------
\put(10,70){\line(0,10){10}}
\put(22.5,74.5){\makebox(0,0)[r]{\small{$a$}}}
%\put(20,88){\vector(0,1){2}}
%\put(20,90){\vector(0,1){2}}
\put(12.5,72.5){\vector(1,1){2.4}}
\put(11.5,71.5){\vector(1,1){2}}
\put(20,76){\vector(0,-1){3}}
%vidurinijo vertikalioji linija
%\put(20,60){\line(0,10){30}}
\put(20,80){\line(-1,-1){8.7}}
\put(20,70){\line(0,10){10}}
\put(20,70){\line(-1,1){8}}
%\put(20,60){\vector(0,1){3}}
%\put(20,58){\vector(0,1){3}}
\put(14.5,75.5){\vector(-1,1){2}}
\put(13.5,76.5){\vector(-1,1){2.4}}
\put(12.2,70.0){\makebox(0,0){\small{$p$}}}
\put(25.5,72){\makebox(0,0){\small{$k^{'}$}}}
\multiput(20,70)(1,0){10}{\circle*{0.35}}
%---------------------------------------
\put(10,60){\line(0,1){10}}
\put(10,60){\vector(0,1){3}}
\put(7.5,63){\makebox(0,0){\small{$n$}}}
\put(10,58){\vector(0,1){3}}
\put(30,60){\line(0,1){20}}
\put(30,80){\vector(0,1){3}}
\put(30,78){\vector(0,1){3}}
\put(27.5,78){\makebox(0,0){\small{$n$}}}
\put(30,60){\vector(0,1){3}}
\put(30,58){\vector(0,1){3}}
\put(27.5,63){\makebox(0,0){\small{$m$}}}
\put(20,53){\makebox(0,0){$\text{A}_{1}$}}
% Second Feynman diagram +30+8
\put(39,74){\makebox(0,0){\normalsize{\bf{+}}}}
\put(48,80){\line(0,1){10}}
\put(48,88){\vector(0,1){2}}
\put(48,90){\vector(0,1){2}}
\put(45.5,88){\makebox(0,0)[t]{\small{$m$}}}
\put(49.9,76.1){\makebox(0,0)[t]{\small{$n$}}}
\put(53,82){\makebox(0,0){\small{$k$}}}
\multiput(48,80)(1,0){10}{\circle*{0.35}}
%---------------------------------------
\put(48,70){\line(0,10){10}}
\put(60.5,74.5){\makebox(0,0)[r]{\small{$a$}}}
%\put(58,88){\vector(0,1){2}}
%\put(58,90){\vector(0,1){2}}
\put(50.5,72.5){\vector(1,1){2.4}}
\put(49.5,71.5){\vector(1,1){2}}
\put(58,76){\vector(0,-1){3}}
%vidurinijo vertikalioji linija
%\put(58,60){\line(0,10){30}}
\put(58,80){\line(-1,-1){8.7}}
\put(58,70){\line(0,10){10}}
\put(58,70){\line(-1,1){8}}
%\put(58,60){\vector(0,1){3}}
%\put(58,58){\vector(0,1){3}}
\put(52.5,75.5){\vector(-1,1){2}}
\put(51.5,76.5){\vector(-1,1){2.4}}
\put(50.2,70.0){\makebox(0,0){\small{$n$}}}
\put(63.5,72){\makebox(0,0){\small{$k^{'}$}}}
\multiput(58,70)(1,0){10}{\circle*{0.35}}
%---------------------------------------
\put(48,60){\line(0,1){10}}
\put(48,60){\vector(0,1){3}}
\put(45.5,63){\makebox(0,0){\small{$p$}}}
\put(48,58){\vector(0,1){3}}
\put(68,60){\line(0,1){20}}
\put(68,76){\vector(0,1){3}}
\put(68,78){\vector(0,1){3}}
\put(65.5,78){\makebox(0,0){\small{$p$}}}
\put(68,60){\vector(0,1){3}}
\put(68,58){\vector(0,1){3}}
\put(65.5,63){\makebox(0,0){\small{$m$}}}
\put(58,53){\makebox(0,0){$\text{A}_{2}$}}
% Third Feynman diagram +30+16
\put(77,74){\makebox(0,0){\normalsize{\bf{+}}}}
\put(86,80){\line(0,1){10}}
\put(86,88){\vector(0,1){2}}
\put(86,90){\vector(0,1){2}}
\put(83.5,88){\makebox(0,0)[t]{\small{$m$}}}
\put(87.9,76.1){\makebox(0,0)[t]{\small{$p$}}}
\put(91,82){\makebox(0,0){\small{$k$}}}
\multiput(86,80)(1,0){10}{\circle*{0.35}}
%---------------------------------------
\put(86,70){\line(0,10){10}}
\put(98.5,74.5){\makebox(0,0)[r]{\small{$a$}}}
%\put(96,88){\vector(0,1){2}}
%\put(96,90){\vector(0,1){2}}
\put(88.5,72.5){\vector(1,1){2.4}}
\put(87.5,71.5){\vector(1,1){2}}
\put(96,76){\vector(0,-1){3}}
%vidurinijo vertikalioji linija
%\put(96,60){\line(0,10){30}}
\put(96,80){\line(-1,-1){8.7}}
\put(96,70){\line(0,10){10}}
\put(96,70){\line(-1,1){8}}
%\put(96,60){\vector(0,1){3}}
%\put(96,58){\vector(0,1){3}}
\put(90.5,75.5){\vector(-1,1){2}}
\put(89.5,76.5){\vector(-1,1){2.4}}
\put(88.2,70.0){\makebox(0,0){\small{$n$}}}
\put(101.5,72){\makebox(0,0){\small{$k^{'}$}}}
\multiput(96,70)(1,0){10}{\circle*{0.35}}
%---------------------------------------
\put(86,60){\line(0,1){10}}
\put(86,60){\vector(0,1){3}}
\put(83.5,63){\makebox(0,0){\small{$n$}}}
\put(86,58){\vector(0,1){3}}
\put(106,60){\line(0,1){20}}
\put(106,80){\vector(0,1){3}}
\put(106,78){\vector(0,1){3}}
\put(103.5,78){\makebox(0,0){\small{$p$}}}
\put(106,60){\vector(0,1){3}}
\put(106,58){\vector(0,1){3}}
\put(103.5,63){\makebox(0,0){\small{$m$}}}
\put(96,53){\makebox(0,0){$\text{A}_{3}$}}
% Fourth Feynman diagram +30+24
\put(115,74){\makebox(0,0){\normalsize{\bf{+}}}}
\put(124,80){\line(0,1){10}}
\put(124,88){\vector(0,1){2}}
\put(124,90){\vector(0,1){2}}
\put(121.5,88){\makebox(0,0)[t]{\small{$m$}}}
\put(125.9,76.1){\makebox(0,0)[t]{\small{$n$}}}
\put(129,82){\makebox(0,0){\small{$k$}}}
\multiput(124,80)(1,0){10}{\circle*{0.35}}
%---------------------------------------
\put(124,70){\line(0,10){10}}
\put(136.5,74.5){\makebox(0,0)[r]{\small{$a$}}}
%\put(134,88){\vector(0,1){2}}
%\put(134,90){\vector(0,1){2}}
\put(126.5,72.5){\vector(1,1){2.4}}
\put(125.5,71.5){\vector(1,1){2}}
\put(134,76){\vector(0,-1){3}}
%vidurinijo vertikalioji linija
%\put(134,60){\line(0,10){30}}
\put(134,80){\line(-1,-1){8.7}}
\put(134,70){\line(0,10){10}}
\put(134,70){\line(-1,1){8}}
%\put(134,60){\vector(0,1){3}}
%\put(134,58){\vector(0,1){3}}
\put(128.5,75.5){\vector(-1,1){2}}
\put(127.5,76.5){\vector(-1,1){2.4}}
\put(126.2,70.0){\makebox(0,0){\small{$p$}}}
\put(139.5,72){\makebox(0,0){\small{$k^{'}$}}}
\multiput(134,70)(1,0){10}{\circle*{0.35}}
%---------------------------------------
\put(124,60){\line(0,1){10}}
\put(124,60){\vector(0,1){3}}
\put(121.5,63){\makebox(0,0){\small{$p$}}}
\put(124,58){\vector(0,1){3}}
\put(144,60){\line(0,1){20}}
\put(144,80){\vector(0,1){3}}
\put(144,78){\vector(0,1){3}}
\put(141.5,78){\makebox(0,0){\small{$n$}}}
\put(144,60){\vector(0,1){3}}
\put(144,58){\vector(0,1){3}}
\put(141.5,63){\makebox(0,0){\small{$m$}}}
\put(134,53){\makebox(0,0){$\text{A}_{4}$}}
%\put(125,64){\makebox(0,0){\normalsize{\bf{=}}}}
% line two
% First Feynman diagram
\put(10,24){\makebox(0,0){\normalsize{\bf{=}}}}
\put(15,24){\makebox(0,0){\LARGE{\Bigg(}}}
\put(20,30){\line(0,1){10}}
\put(20,38){\vector(0,1){2}}
\put(20,40){\vector(0,1){2}}
\put(17.5,38){\makebox(0,0)[t]{\small{$m$}}}
\put(21.9,26.1){\makebox(0,0)[t]{\small{$p$}}}
\put(25,32){\makebox(0,0){\small{$k$}}}
\multiput(20,30)(1,0){10}{\circle*{0.35}}
%---------------------------------------
\put(20,20){\line(0,10){10}}
\put(32.5,24.5){\makebox(0,0)[r]{\small{$a$}}}
%\put(30,38){\vector(0,1){2}}
%\put(30,40){\vector(0,1){2}}
\put(22.5,22.5){\vector(1,1){2.4}}
\put(21.5,21.5){\vector(1,1){2}}
\put(30,26){\vector(0,-1){3}}
%vidurinijo vertikalioji linija
%\put(30,10){\line(0,10){30}}
\put(30,30){\line(-1,-1){8.7}}
\put(30,20){\line(0,10){10}}
\put(30,20){\line(-1,1){8}}
%\put(30,10){\vector(0,1){3}}
%\put(30,08){\vector(0,1){3}}
\put(24.5,25.5){\vector(-1,1){2}}
\put(23.5,26.5){\vector(-1,1){2.4}}
\put(22.2,20.0){\makebox(0,0){\small{$p$}}}
\put(35.5,22){\makebox(0,0){\small{$k^{'}$}}}
\multiput(30,20)(1,0){10}{\circle*{0.35}}
%---------------------------------------
\put(20,10){\line(0,1){10}}
\put(20,10){\vector(0,1){3}}
\put(17.5,13){\makebox(0,0){\small{$n$}}}
\put(20,08){\vector(0,1){3}}
\put(40,10){\line(0,1){20}}
\put(40,30){\vector(0,1){3}}
\put(40,28){\vector(0,1){3}}
\put(37.5,28){\makebox(0,0){\small{$n$}}}
\put(40,10){\vector(0,1){3}}
\put(40,08){\vector(0,1){3}}
\put(37.5,13){\makebox(0,0){\small{$m$}}}
\put(30,03){\makebox(0,0){$\text{A}_{5}$}}
% Second Feynman diagram +30
\put(49,24){\makebox(0,0){\normalsize{\bf{+}}}}
\put(58,30){\line(0,1){10}}
\put(58,38){\vector(0,1){2}}
\put(58,40){\vector(0,1){2}}
\put(55.5,38){\makebox(0,0)[t]{\small{$m$}}}
\put(59.9,26.1){\makebox(0,0)[t]{\small{$p$}}}
\put(63,32){\makebox(0,0){\small{$k$}}}
\multiput(58,30)(1,0){10}{\circle*{0.35}}
%---------------------------------------
\put(58,20){\line(0,10){10}}
\put(70.6,24.5){\makebox(0,0)[r]{\small{$a$}}}
%\put(68,38){\vector(0,1){2}}
%\put(68,40){\vector(0,1){2}}
\put(60.5,22.5){\vector(1,1){2.4}}
\put(59.5,21.5){\vector(1,1){2}}
\put(68,26){\vector(0,-1){3}}
%vidurinijo vertikalioji linija
%\put(68,10){\line(0,10){30}}
\put(68,30){\line(-1,-1){8.7}}
\put(68,20){\line(0,10){10}}
\put(68,20){\line(-1,1){8}}
%\put(68,10){\vector(0,1){3}}
%\put(68,08){\vector(0,1){3}}
\put(62.5,25.5){\vector(-1,1){2}}
\put(61.5,26.5){\vector(-1,1){2.4}}
\put(60.2,20.0){\makebox(0,0){\small{$n$}}}
\put(73.5,22){\makebox(0,0){\small{$k^{'}$}}}
\multiput(68,20)(1,0){10}{\circle*{0.35}}
%---------------------------------------
\put(58,10){\line(0,1){10}}
\put(58,10){\vector(0,1){3}}
\put(55.5,13){\makebox(0,0){\small{$n$}}}
\put(58,08){\vector(0,1){3}}
\put(78,10){\line(0,1){20}}
\put(78,30){\vector(0,1){3}}
\put(78,28){\vector(0,1){3}}
\put(75.5,28){\makebox(0,0){\small{$p$}}}
\put(78,10){\vector(0,1){3}}
\put(78,08){\vector(0,1){3}}
\put(75.5,13){\makebox(0,0){\small{$m$}}}
\put(68,03){\makebox(0,0){$\text{A}_{6}$}}
\put(83,24){\makebox(0,0){\LARGE{\Bigg)}}}
%\put(100,24){\makebox(0,0){\LARGE{\Bigg( 1 + P \big( $n \rightleftharpoons p$ \big)\Bigg)}}}
\put(109,24){\makebox(0,0){\LARGE{\Bigg( 1 + P $
 \left( %\hspace{-0.15cm}
\begin{array}{lcl}
      n&\hspace{-0.25cm}\rightleftharpoons&\hspace{-0.25cm}p \\
%      k&\hspace{-0.25cm}\rightleftharpoons&\hspace{-0.25cm}k^{'}
    \end{array}  \hspace{-0.15cm} \right)
$ \Bigg)}}}
\end{picture}
\caption{The CV Feynman diagrams of the second-order effective Hamiltonian, expressing the fourth type of core-valence correlations
$(n_{a} \ell_{a})\, j_{a}^{2j_a+1} \, (n_{m} \ell_{m})\, j_{m}^{w_m} \, (n_{n} \ell_{n})\, j_{n}^{w_n} \, (n_{p} \ell_{p})\, j_{p}^{w_p} 
   \rightarrow (n_{a} \ell_{a})\, j_{a}^{2j_a} \; (n_{m} \ell_{m})\, j_{m}^{w_m-1} \, (n_{n} \ell_{n})\,j_{n}^{w_n+1}  \; (n_{p} \ell_{p})\, j_{p}^{w_p+1}$.
%$(n_{m} \ell_{m})\, j_{m}^{w_m} \, (n_{n} \ell_{n})\, j_{n}^{w_n} \, (n_{p} \ell_{p})\, j_{p}^{w_p} 
%   \rightarrow (n_{m} \ell_{m})\, j_{m}^{w_m+1} \; (n_{n} \ell_{n})\, j_{n}^{w_n-1} \; (n_{p} \ell_{p})\, j_{p}^{w_p-1} \; (n_{r} \ell_{r})\, j_{r}$.
}
\label{CV_4_type}
\end{center}
\end{figure*}
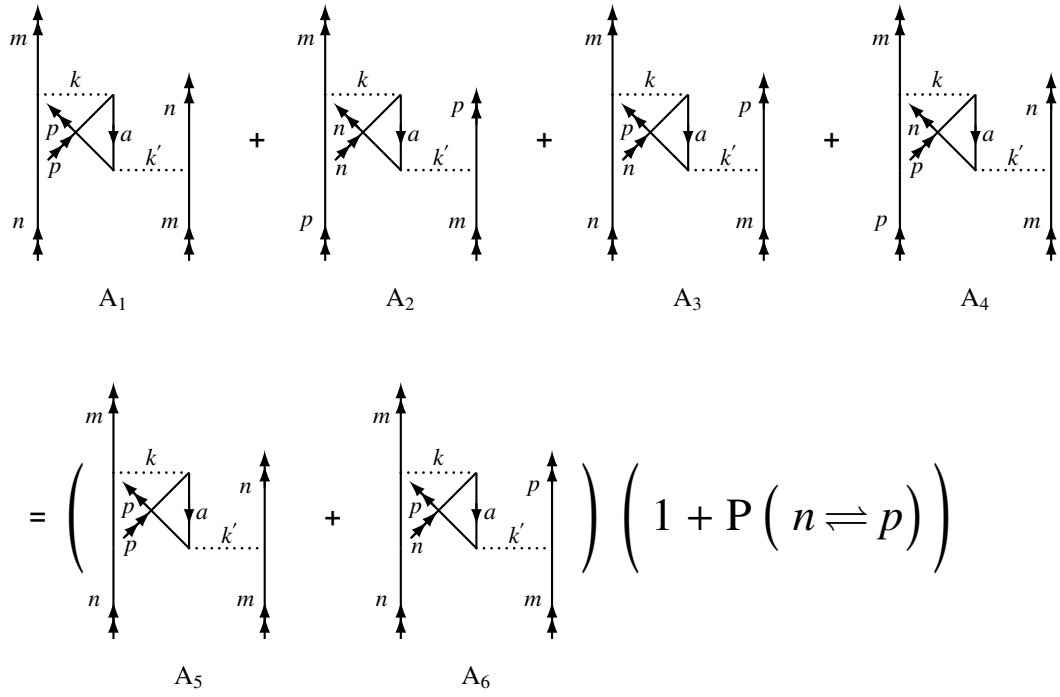

The following type of correlations
\begin{equation}
\label{eq:CVT-b}
    (n_{a} \ell_{a})\, j_{a}^{2j_a+1} \, (n_{m} \ell_{m})\, j_{m}^{w_m} \, (n_{n} \ell_{n})\, j_{n}^{w_n} \, (n_{p} \ell_{p})\, j_{p}^{w_p} 
% \nonumber \\
   \rightarrow (n_{a} \ell_{a})\, j_{a}^{2j_a} \; (n_{m} \ell_{m})\, j_{m}^{w_m-1} \, (n_{n} \ell_{n})\,j_{n}^{w_n+1}  \; (n_{p} \ell_{p})\, j_{p}^{w_p+1}
\end{equation}
is described by the same diagram CV$_7$ with four different sets of open lines with double-arrow indices. 
Four diagrams A$_1$, A$_2$, A$_3$, and A$_4$ from Fig.~\ref{CV_4_type},
 represent all these different sets.
For example, Feynman diagram A$_1$ have the following values $m=p' \equiv m$, \, $m'=p \equiv n$, and $n=n' \equiv p$.
Therefore, to find the value of this type of correlations, it is necessary to analyze all four diagrams: A$_1$, A$_2$, A$_3$, and A$_4$. 
But it is important to note that, to achieve the same result, the four diagrams can be easily replaced by just two diagrams, 
A$_5$ (A$_5 \equiv$ A$_1$), A$_6$ (A$_6 \equiv$ A$_3$), plus a multiplier $\bigg( 1 + P \left( ... \right) \bigg)$, where the notation 
$P \left(  \hspace{-0.15cm}
\begin{array}{lcl}
      n&\hspace{-0.25cm}\rightleftharpoons&\hspace{-0.25cm}p \\
%      k&\hspace{-0.25cm}\rightleftharpoons&\hspace{-0.25cm}k^{'}
    \end{array}  \hspace{-0.15cm} \right)$
means that the indices $n$ and $p$ should be swapped.
This makes it considerably easier to carry out the desired calculations. 
Note that A$_5$ describes the direct part of the correlation under consideration, while A$_6$ describes the exchange part of the interaction.
These A$_5$ and A$_6$ are three-particle Feynman diagrams for which the program library \texttt{librang}~\cite{Gaigalas:2022} from {\sc Grasp}
does not support the calculation of spin-angular parts. It is therefore necessary to extend the capabilities of this program library to calculate the values of these diagrams. This problem will be discussed and solved in detail in the next section.

\section{The spin-angular part of the three-particle Feynman diagram contributing to core-valence correlations}
\label{Sec:spin_angular}

The program library \texttt{librang}~\cite{Gaigalas:2022} from {\sc Grasp} evaluates spin-angular coefficients of any matrix element with any number of open subshells for any one- and/or two-particle operators. 
Therefore, in previous papers~\cite{Gaigetal:2024CV,Gaigetal:2024C,Gaigetal:2024CC,Gaigetal:2025VV}, when developing the combination of
second-order Rayleigh-Schr\"odinger perturbation theory
and relativistic configuration interaction method, there was no problem in using it to find correlations described by vacuum, one- and two-particle Feynman diagrams.
However, problems with using this library arise when dealing with three-particle operators~\cite{Gaigetal:2025VVT}.

The program library \texttt{librang}~\cite{Gaigalas:2022} is based on the spin-angular approach~\cite{Gaigalas_1996,Gaigalas_1997,Gaigalas:2026a}. The specifics of this approach (factorization of standard values, interaction strengths, and recoupling matrices) make it easy enough to extend it to allow the program library to find the spin-angular part of the CV$_7$ Feynman diagram that is needed to find the correlations that are being studied in this paper. This can be done by using ideas published in the papers~\cite{Gaigalas:85,Gaigalas:89}, where the second quantization operators are grouped according to their action on the subshell, i.e., so that the second quantization operators acting on the subshell $m$ first, then on the subshell $p$ (if there are any), and finally on the subshell $n$. The spin-angular library extension, both in the paper~\cite{Gaigetal:2025VVT} and in this paper, also uses quasispin symmetry and formalism~\cite{Gaigalas_1996,Gaigalas_1997,Gaigalas:2026a}, as in the original library \texttt{librang}~\cite{Gaigalas:2022}. This greatly facilitates spin-angular integration. Below, we show how this has been done for the different types of core-valence correlations described by a three-particle Feynman diagram, separately.

\subsection{The spin-angular part of the third type of core-valence correlations}
\label{sec:PT_SA_Third}

In this case, the diagram CV$_7$ has the following tensorial structure
\begin{equation}
\label{eq:Tensor11}
\biggl[ \Bigl[ \bigl[ a^{(j_m)}_1 \times \tilde{a}^{(j_n)}_2 \bigr]^{(k)}   \times \bigl[ \tilde{a}^{(j_n)}_3 \times a^{(j_n)}_4 \bigr]^{(x)} \Bigr]^{(k')}  \times \bigl[ a^{(j_n)}_5 \times \tilde{a}^{(j_m)}_6 \bigr]^{(k')} \biggr]^{(0)},
\end{equation}
where the subscript next to the operator indicates the sequence number of the second quantization operator in the tensorial product.
As seen in the tensorial structure (\ref{eq:Tensor11}), the two pairs of second quantization operators are mixed, i.e., they consist of second quantization operators acting on different subshells. This is a pair consisting of operators with sequence numbers 1 and 2 and a pair with sequence numbers 5 and 6. This situation greatly complicates the study of the spin-angular part of this operator. But if we replace the tensorial structure (\ref{eq:Tensor11}) with the following
\begin{equation}
\label{eq:Tensor12}
\biggl[ \bigl[ a^{(j_m)}_1 \times \tilde{a}^{(j_m)}_6 \bigr]^{(J_1)} \times \Bigl[ \bigl[ a^{(j_n)}_4 \times \tilde{a}^{(j_n)}_3 \bigr]^{(x)} \times \bigl[ \tilde{a}^{(j_n)}_2 \times a^{(j_n)}_5 \bigr]^{(J_2)}  \Bigr]^{(J_1)} \biggr]^{(0)},
\end{equation}
then the situation would change and one could easily extend the spin-angular approach~\cite{Gaigalas_1996,Gaigalas_1997,Gaigalas:2026a}
to the study of three-particle operator for this type of correlations~\cite{Gaigalas:89} as was done in~\cite{Gaigetal:2025VVT}.
We also point out that this allows us to use all known symmetries of the atom, including the quasispin symmetry~\cite{Gaigalas_1996,Gaigalas_1997}, in the calculations.

There are several such options for the tensorial product. But we have chosen the form (\ref{eq:Tensor12}) because the transformation to (\ref{eq:Tensor12}) is similar to the transformation from expression (13) to (14) in~\cite{Gaigetal:2025VVT}, and it is possible to make use of the recoupling coefficients of \cite{Gaigetal:2025VVT} represented by diagrams B$_1$ and B$_2$. But first of all, using the well-known commutation rule of secondary quantization (see, for example, (29) from~\cite{Judd:67}, where expression is in $LS$-coupling) 
\begin{equation}
\label{eq:Tensor13}
\bigl[ \tilde{a}^{(j_i)} \times a^{(j_k)} \bigr]^{(x)} =
 -\left( -1 \right)^{j_i+j_k-x} \bigl[ a^{(j_k)}  \times \tilde{a}^{(j_i)} \bigr]^{(x)}
+ \sqrt{\left[ j_i\right]} \; \delta \left( n_i l_i j_i, n_k l_k j_k \right) \; \delta \left( x, 0 \right),
\end{equation}
we interchange the operators $a_3$ and $a_4$. This leads to the expression
\begin{equation}
\label{eq:Tensor14}
\begin{split}
\biggl[ \Bigl[ \bigl[ a^{(j_m)}_1 \times \tilde{a}^{(j_n)}_2 \bigr]^{(k)}   \times \bigl[ \tilde{a}^{(j_n)}_3 \times a^{(j_n)}_4 \bigr]^{(x)} \Bigr]^{(k')}  \times \bigl[ a^{(j_n)}_5 \times \tilde{a}^{(j_m)}_6 \bigr]^{(k')} \biggr]^{(0)}
\hspace{2.cm}\\
 = \left( -1 \right)^{x} 
\biggl[ \Bigl[ \bigl[ a^{(j_m)}_1 \times \tilde{a}^{(j_n)}_2 \bigr]^{(k)}   \times \bigl[ a^{(j_n)}_4 \times \tilde{a}^{(j_n)}_3 \bigr]^{(x)} \Bigr]^{(k')}  \times \bigl[ a^{(j_n)}_5 \times \tilde{a}^{(j_m)}_6 \bigr]^{(k')} \biggr]^{(0)}
 \\
 + \sqrt{\left[ j_n\right]} \; \Bigl[ \bigl[ a^{(j_m)}_1 \times \tilde{a}^{(j_n)}_2 \bigr]^{(k)} \times \bigl[ a^{(j_n)}_5 \times \tilde{a}^{(j_m)}_6 \bigr]^{(k')} \Bigr]^{(0)} \delta \left( x,0 \right) \; \delta \left( k,k' \right).
\end{split}
\end{equation}
Now, in the first part of expression (\ref{eq:Tensor14}), using the same commutation rule (\ref{eq:Tensor13}), after interchanging the operators of the first and third pair of the operators of secondary quantization, we get
\begin{equation}
\label{eq:Tensor15}
\begin{split}
\biggl[ \Bigl[ \bigl[ a^{(j_m)}_1 \times \tilde{a}^{(j_n)}_2 \bigr]^{(k)} \times \bigl[ \tilde{a}^{(j_n)}_3 \times a^{(j_n)}_4 \bigr]^{(x)} \Bigr]^{(k')}  \times \bigl[ a^{(j_n)}_5 \times \tilde{a}^{(j_m)}_6 \bigr]^{(k')} \biggr]^{(0)}
\hspace{2.cm}\\
 = \left( -1 \right)^{x+k+k'} 
\biggl[ \Bigl[ \bigl[\tilde{a}^{(j_n)}_2 \times a^{(j_m)}_1 \bigr]^{(k)} \times \bigl[ a^{(j_n)}_4 \times \tilde{a}^{(j_n)}_3 \bigr]^{(x)} \Bigr]^{(k')}  \times \bigl[ \tilde{a}^{(j_m)}_6 \times a^{(j_n)}_5 \bigr]^{(k')} \biggr]^{(0)}
 \\
 + \sqrt{\left[ j_n\right]} \; \Bigl[ \bigl[ a^{(j_m)}_1 \times \tilde{a}^{(j_n)}_2 \bigr]^{(k)} \times \bigl[ a^{(j_n)}_5 \times \tilde{a}^{(j_m)}_6 \bigr]^{(k')} \Bigr]^{(0)} \delta \left( x,0 \right) \; \delta \left( k,k' \right).
\end{split}
\end{equation}
Now, similarly to the case of~\cite{Gaigetal:2025VVT} (7), using the commutation rules of the operators of second quantization and the graphical representation~\cite{Gaigatal:85} of the angular momentum theory, we transform the first part of the expression (\ref{eq:Tensor15}) into the tensorial structure (\ref{eq:Tensor12}), and the second part of the expression we transform in such way that the pair of operators of second quantization act on the same subshell. Thus, we get:
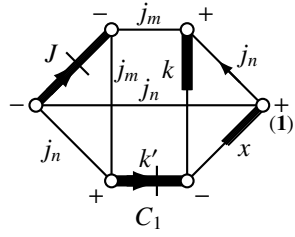
\begin{figure}
\begin{center}
\setlength{\unitlength}{1mm}
\begin{picture}(47,37)
\thicklines
%---------------------------------------
\put(12,27){\makebox(0,0)[t]{$J$}}
\put(38,27){\makebox(0,0)[t]{$j_{n}$}}
\put(22,25){\makebox(0,0)[t]{$j_{m}$}}
\put(27.5,25){\makebox(0,0)[t]{$k$}}
%vir
\put(18.5,31){\makebox(0,0){$-$}}
\put(32.5,31){\makebox(0,0){$+$}}
%\put(30,31.5){\makebox(0,0){\footnotesize{\bf{2}}}}
\put(20,29){\line(10,0){10}}
\put(25,31){\makebox(0,0){$j_{m}$}}
%v
\put(10,19){\line(30,0){30}}
%\put(33,26){\vector(-1,1){3}}
\put(37,22){\vector(-1,1){3}}
\put(25,21){\makebox(0,0){$j_{n}$}}
%ap
\put(18,07){\makebox(0,0){$+$}}
%\put(20,6.5){\makebox(0,0){\footnotesize{\bf{1}}}}
\put(32,07){\makebox(0,0){$-$}}
\put(20,09.6){\line(10,0){10}}
\put(20,09.5){\line(10,0){10}}
\put(20,09.4){\line(10,0){10}}
\put(20,09.3){\line(10,0){10}}
\put(20,09.2){\line(10,0){10}}
\put(20,09.1){\line(10,0){10}}
\put(20,09){\line(10,0){10}}
\put(20,08.9){\line(10,0){10}}
\put(20,08.8){\line(10,0){10}}
\put(20,08.7){\line(10,0){10}}
\put(20,08.6){\line(10,0){10}}
\put(20,08.5){\line(10,0){10}}
\put(20,08.4){\line(10,0){10}}
\put(26,07){\line(0,4){4}}
\put(22,09.6){\vector(1,0){3}}
\put(22,08.4){\vector(1,0){3}}
\put(25,12){\makebox(0,0){$k'$}}
%---------------------------------------
\put(9.4,19){\line(1,1){10}}
\put(9.7,19){\line(1,1){10}}
\put(10,19){\line(1,1){10}}
\put(10.3,19){\line(1,1){10}}
\put(10.6,19){\line(1,1){10}}
%GG\put(14,20){\line(-2,2){3}}
%GG\put(18.4,28.0){\vector(-1,-1){3}}
%GG\put(19,27.4){\vector(-1,-1){3}}
\put(17,23){\line(-2,2){3}}
\put(11.4,21.0){\vector(1,1){3}}
\put(12,20.4){\vector(1,1){3}}
\put(10,19){\line(1,-1){10}}
\put(39.4,19){\line(-1,-1){4.8}}
\put(39.7,19){\line(-1,-1){4.9}}
\put(40,19){\line(-1,1){10}}
\put(40,19){\line(-1,-1){10}}
\put(40.3,19){\line(-1,-1){5.2}}
\put(40.7,19){\line(-1,-1){5.3}}
%---------------------------------------
\put(20,09){\line(0,20){20}}
\put(12,13){\makebox(0,0){$j_{n}$}}
\put(07.5,19){\makebox(0,0){$-$}}
\put(29.4,29){\line(0,-8){8}}
\put(29.5,29){\line(0,-8){8}}
\put(29.6,29){\line(0,-8){8}}
\put(29.7,29){\line(0,-8){8}}
\put(29.8,29){\line(0,-8){8}}
\put(29.9,29){\line(0,-8){8}}
\put(30,09){\line(0,20){20}}
\put(30.1,29){\line(0,-8){8}}
\put(30.2,29){\line(0,-8){8}}
\put(30.3,29){\line(0,-8){8}}
\put(30.4,29){\line(0,-8){8}}
\put(30.5,29){\line(0,-8){8}}
\put(30.6,29){\line(0,-8){8}}
\put(37.5,13){\makebox(0,0){$x$}}
\put(42.5,19){\makebox(0,0){$+$}}
%\put(44.5,19){\makebox(0,0){\footnotesize{\bf{'1'}}}}
\put(42.5,16.4){\makebox(0,0){\footnotesize{({\bf 1})}}}
\put(25,04){\makebox(0,0){$ C_{1}$}}
%--------------------------------------- tusciaviduriai mazgai
%\put(20,29){\circle*{1.7}}
\put(20,29){\color{white}\circle*{1.7}} % Fill white
\put(20,29){\color{black}\circle{1.7}}  % Outline
%\put(30,29){\circle*{1.7}}
\put(30,29){\color{white}\circle*{1.7}} % Fill white
\put(30,29){\color{black}\circle{1.7}}  % Outline
%\put(10,19){\circle*{1.7}}
\put(10,19){\color{white}\circle*{1.7}} % Fill white
\put(10,19){\color{black}\circle{1.7}}  % Outline
%\put(40,19){\circle*{1.7}}
\put(40,19){\color{white}\circle*{1.7}} % Fill white
\put(40,19){\color{black}\circle{1.7}}  % Outline
%\put(20,09){\circle*{1.7}}
\put(20,09){\color{white}\circle*{1.7}} % Fill white
\put(20,09){\color{black}\circle{1.7}}  % Outline
%\put(30,09){\circle*{1.7}}
\put(30,09){\color{white}\circle*{1.7}} % Fill white
\put(30,09){\color{black}\circle{1.7}}  % Outline
\end{picture}
\caption{Diagram showing the recoupling coefficient $C_1$.}
\label{P6-B1}
\end{center}
\end{figure}
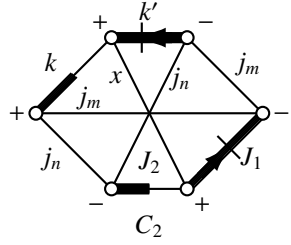
\begin{figure}
\begin{center}
\setlength{\unitlength}{1mm}
\begin{picture}(47,37)
\thicklines
%---------------------------------------
\put(12,27){\makebox(0,0)[t]{$k$}}
\put(38,27){\makebox(0,0)[t]{$j_{m}$}}
\put(20.5,24.5){\makebox(0,0)[t]{$x$}}
\put(29,25){\makebox(0,0)[t]{$j_n$}}
%vir
\put(18.5,31){\makebox(0,0){$+$}}
\put(32.5,31){\makebox(0,0){$-$}}
%\put(30,31.5){\makebox(0,0){\footnotesize{\bf{2}}}}
%\put(35,32){\makebox(0,0){2}}
\put(20,29.6){\line(10,0){10}}
\put(20,29.5){\line(10,0){10}}
\put(20,29.4){\line(10,0){10}}
\put(20,29.3){\line(10,0){10}}
\put(20,29.2){\line(10,0){10}}
\put(20,29.1){\line(10,0){10}}
\put(20,29){\line(10,0){10}}
\put(20,28.9){\line(10,0){10}}
\put(20,28.8){\line(10,0){10}}
\put(20,28.7){\line(10,0){10}}
\put(20,28.6){\line(10,0){10}}
\put(20,28.5){\line(10,0){10}}
\put(20,28.4){\line(10,0){10}}
\put(24,27){\line(0,4){4}}
\put(28,29.6){\vector(-1,0){3}}
\put(28,28.4){\vector(-1,0){3}}
\put(25,32.5){\makebox(0,0){$k'$}}
%v
\put(10.5,19){\line(29,0){30}}
\put(17,21){\makebox(0,0){$j_{m}$}}
%ap
\put(18,07){\makebox(0,0){$-$}}
\put(32,07){\makebox(0,0){$+$}}
\put(20,09.6){\line(5,0){5}}
\put(20,09.5){\line(5,0){5}}
\put(20,09.4){\line(5,0){5}}
\put(20,09.3){\line(5,0){5}}
\put(20,09.2){\line(5,0){5}}
\put(20,09.1){\line(5,0){5}}
\put(20,09){\line(10,0){10}}
\put(20,08.9){\line(5,0){5}}
\put(20,08.8){\line(5,0){5}}
\put(20,08.7){\line(5,0){5}}
\put(20,08.6){\line(5,0){5}}
\put(20,08.5){\line(5,0){5}}
\put(20,08.4){\line(5,0){5}}
%\put(26,07){\line(0,4){4}}
%\put(22,09.6){\vector(1,0){3}}
%\put(22,08.4){\vector(1,0){3}}
\put(25,12.5){\makebox(0,0){$J_{2}$}}
%---------------------------------------
\put(9.4,19){\line(1,1){5.3}}
\put(9.7,19){\line(1,1){5.2}}
\put(10,19){\line(1,1){10}}
\put(10.3,19){\line(1,1){4.9}}
\put(10.6,19){\line(1,1){4.8}}
\put(10,19){\line(1,-1){10}}
\put(40,19){\line(-1,1){10}}
\put(39.4,19){\line(-1,-1){10}}
\put(39.7,19){\line(-1,-1){10}}
\put(40,19){\line(-1,-1){10}}
\put(40.3,19){\line(-1,-1){10}}
\put(40.7,19){\line(-1,-1){10}}
%\put(34,10){\line(-2,2){3}}
\put(37,13){\line(-2,2){3}}
\put(31.4,11.0){\vector(1,1){3}}
\put(32,10.4){\vector(1,1){3}}
%---------------------------------------
\put(20,09){\line(1,2){10}}
\put(12,13){\makebox(0,0){$j_{n}$}}
\put(07.5,19){\makebox(0,0){$+$}}
%\put(05.5,19){\makebox(0,0){\footnotesize{\bf{1}}}}
%\put(05.5,19){\makebox(0,0){1}}
\put(30,09){\line(-1,2){10}}
\put(38.5,13){\makebox(0,0){$J_{1}$}}
\put(42.5,19){\makebox(0,0){$-$}}
\put(25,04){\makebox(0,0){$ C_{2}$}}
%--------------------------------------- tusciaviduriai mazgai
%\put(20,29){\circle*{1.7}}
\put(20,29){\color{white}\circle*{1.7}} % Fill white
\put(20,29){\color{black}\circle{1.7}}  % Outline
%\put(30,29){\circle*{1.7}}
\put(30,29){\color{white}\circle*{1.7}} % Fill white
\put(30,29){\color{black}\circle{1.7}}  % Outline
%\put(10,19){\circle*{1.7}}
\put(10,19){\color{white}\circle*{1.7}} % Fill white
\put(10,19){\color{black}\circle{1.7}}  % Outline
%\put(40,19){\circle*{1.7}}
\put(40,19){\color{white}\circle*{1.7}} % Fill white
\put(40,19){\color{black}\circle{1.7}}  % Outline
%\put(20,09){\circle*{1.7}}
\put(20,09){\color{white}\circle*{1.7}} % Fill white
\put(20,09){\color{black}\circle{1.7}}  % Outline
%\put(30,09){\circle*{1.7}}
\put(30,09){\color{white}\circle*{1.7}} % Fill white
\put(30,09){\color{black}\circle{1.7}}  % Outline
\end{picture}
\caption{Diagram showing the recoupling coefficient $C_2$.}
\label{P6_B2}
\end{center}
\end{figure}
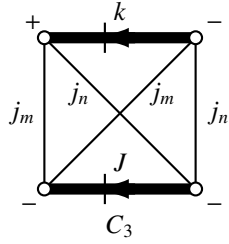
\begin{figure}
\begin{center}
\setlength{\unitlength}{1mm}
\begin{picture}(40,37)
\thicklines
%---------------------------------------
\put(25.6,23){\makebox(0,0)[t]{$j_{m}$}}
\put(14.5,23){\makebox(0,0)[t]{$j_n$}}
%vir
\put(08.5,31){\makebox(0,0){$+$}}
\put(32.5,31){\makebox(0,0){$-$}}
\put(10,29.6){\line(20,0){20}}
\put(10,29.5){\line(20,0){20}}
\put(10,29.4){\line(20,0){20}}
\put(10,29.3){\line(20,0){20}}
\put(10,29.2){\line(20,0){20}}
\put(10,29.1){\line(20,0){20}}
\put(10,29){\line(20,0){20}}
\put(10,28.9){\line(20,0){20}}
\put(10,28.8){\line(20,0){20}}
\put(10,28.7){\line(20,0){20}}
\put(10,28.6){\line(20,0){20}}
\put(10,28.5){\line(20,0){20}}
\put(10,28.4){\line(20,0){20}}
\put(18,27){\line(0,4){4}}
\put(22,29.6){\vector(-1,0){3}}
\put(22,28.4){\vector(-1,0){3}}
\put(20,32.5){\makebox(0,0){$k$}}

%v
\put(07,19){\makebox(0,0){$j_{m}$}}
%ap
\put(08,07){\makebox(0,0){$-$}}
\put(32.5,07){\makebox(0,0){$-$}}
\put(10,09.6){\line(20,0){20}}
\put(10,09.5){\line(20,0){20}}
\put(10,09.4){\line(20,0){20}}
\put(10,09.2){\line(20,0){20}}
\put(10,09.3){\line(20,0){20}}
\put(10,09.1){\line(20,0){20}}
\put(10,09){\line(20,0){20}}
\put(10,08.9){\line(20,0){20}}
\put(10,08.8){\line(20,0){20}}
\put(10,08.7){\line(20,0){20}}
\put(10,08.6){\line(20,0){20}}
\put(10,08.5){\line(20,0){20}}
\put(10,08.4){\line(20,0){20}}
\put(18,7){\line(0,4){4}}
\put(22,9.6){\vector(-1,0){3}}
\put(22,8.4){\vector(-1,0){3}}
\put(20,12.5){\makebox(0,0){$J$}}
\put(33,19){\makebox(0,0){$j_{n}$}}
%---------------------------------------
\put(10,09){\line(0,1){20}}
\put(30,09){\line(0,1){20}}
\put(10,09){\line(1,1){20}}
\put(30,09){\line(-1,1){20}}
\put(20,04){\makebox(0,0){$ C_{3}$}}
%--------------------------------------- tusciaviduriai mazgai
%\put(10,29){\circle*{1.7}}
\put(10,29){\color{white}\circle*{1.7}} % Fill white
\put(10,29){\color{black}\circle{1.7}}  % Outline
%\put(30,29){\circle*{1.7}}
\put(30,29){\color{white}\circle*{1.7}} % Fill white
\put(30,29){\color{black}\circle{1.7}}  % Outline
%\put(30,09){\circle*{1.7}}
\put(30,09){\color{white}\circle*{1.7}} % Fill white
\put(30,09){\color{black}\circle{1.7}}  % Outline
%\put(10,09){\circle*{1.7}}
\put(10,09){\color{white}\circle*{1.7}} % Fill white
\put(10,09){\color{black}\circle{1.7}}  % Outline
\end{picture}
\caption{Diagram showing the recoupling coefficient $C_3$.}
\label{B3}
\end{center}
\end{figure}
\begin{equation}
\label{eq:Tensor16}
\begin{split}
\biggl[ \Bigl[ \bigl[ a^{(j_m)}_1 \times \tilde{a}^{(j_n)}_2 \bigr]^{(k)} \times \bigl[ \tilde{a}^{(j_n)}_3 \times a^{(j_n)}_4 \bigr]^{(x)} \Bigr]^{(k')}  \times \bigl[ a^{(j_n)}_5 \times \tilde{a}^{(j_m)}_6 \bigr]^{(k')} \biggr]^{(0)}
\hspace{2.cm}\\
 = \left( -1 \right)^{x+k+k'} C_1 \; \Bigl[ \bigl[ a^{(j_m)}_1 \times \tilde{a}^{(j_m)}_6 \bigr]^{(J)} \times \bigl[ \tilde{a}^{(j_n)}_3 \times  a^{(j_n)}_5  \bigr]^{(J)} \Bigr]^{(0)} 
\\
+ \left( -1 \right)^{x+k+k'} C_2 \;
\biggl[ \bigl[ a^{(j_m)}_1 \times \tilde{a}^{(j_m)}_6 \bigr]^{(J_1)} \times \Bigl[ \bigl[ a^{(j_n)}_4 \times \tilde{a}^{(j_n)}_3 \bigr]^{(x)} \times \bigl[ \tilde{a}^{(j_n)}_2 \times a^{(j_n)}_5 \bigr]^{(J_2)}  \Bigr]^{(J_1)} \biggr]^{(0)}
 \\
 + C_3 \; \sqrt{\left[ j_n\right]} \; \Bigl[ \bigl[ a^{(j_m)}_1 \times \tilde{a}^{(j_m)}_6 \bigr]^{(J)} \times \bigl[  \tilde{a}^{(j_n)}_2 \times a^{(j_n)}_5\bigr]^{(J)} \Bigr]^{(0)} \delta \left( x,0 \right) \; \delta \left( k,k' \right),
\end{split}
\end{equation}
%\gg{kur daugikliai $B_1$, $B_2$ ir $B_3$ yra perrisimo matricos, kurios ateina perrisus tenzorines sandaugas i pavidala pateikta israiskoje (\ref{eq:Tensor16}). Palyginus diagrama $B_1$ is [5] su sio straipsnio $B_1$ diagrama, matome, kad sio straipsnio diagramoje ant linjos $j_n$ yra rodykle priesingos krypties bei prie mazko $({\bf 1})$ yra priesingas zenklas. Visa kita yra tas pats. Todel, norint tureti sio straipsnio diagramos $B_1$ analizine israiska, tereikia pasinaudoti [5] straipsnio $B_1$ diagramos algebrine israiska, ja papildomai padauginus is fazinio daugiklio 
%$\left( -1 \right)^{2j_n} + \left( -1 \right)^{1+2j_n -x} \equiv \left(-1\right)^{1+x}$. $B_2$ diagrama yra identiska $B_2$ diagramai is straipsnio [5], o $B_3$ yra %naujai ateinati diagrama. 
%Pagal the graphical representation~\cite{Gaigatal:85} of the angular momentum theory si diagrama $B_3$ yra lygi 6$j$-symbol su papildomais daugikliais.
%Istacius siu diagramu $B_1$, $B_2$ ir $B_3$ algebrines israiskas, galutinai (\ref{eq:Tensor16}) turi tokia analizine israiska:}
where the multipliers $C_1$, $C_2$ and $C_3$ are the recoupling matrices that come from the recoupling of the tensorial products to the form given in the expression (\ref{eq:Tensor16}). Comparing the diagram $B_1$ from~\cite{Gaigetal:2025VVT} with the diagram $C_1$ of this paper, we see that in the diagram of this paper there is an arrow on the line $j_n$ in the opposite direction and an opposite sign at the node $(1)$. The rest is the same. Therefore, to have an analytical expression for the diagram $C_1$ in this paper, we only need to use the algebraic expression of the diagram $B_1$ in~\cite{Gaigetal:2025VVT}, multiplied by the phase multiplier $\left( -1 \right)^{2j_n} \times \left( -1 \right)^{1+2j_n -x} \equiv \left(-1 \right)^{1+x}$. The $C_2$ diagram is identical to the $B_2$ diagram from paper~\cite{Gaigetal:2025VVT}, i.e., $C_2 \equiv B_2$, and $C_3$ is a new arrival. According to the graphical representation of angular momentum theory~\cite{Gaigatal:85,JucBan:77a}, the diagram $C_3$ is equal to the 6-$j$ symbol with additional multipliers. The algebraic expressions for $C_1$, $C_2$ and $C_3$ of these diagrams lead to the following analytical expression (\ref{eq:Tensor16}):

\begin{equation}
\label{eq:Tensor17}
\begin{split}
\biggl[ \Bigl[ \bigl[ a^{(j_m)}_1 \times \tilde{a}^{(j_n)}_2 \bigr]^{(k)} \times \bigl[ \tilde{a}^{(j_n)}_3 \times a^{(j_n)}_4 \bigr]^{(x)} \Bigr]^{(k')}  \times \bigl[ a^{(j_n)}_5 \times \tilde{a}^{(j_m)}_6 \bigr]^{(k')} \biggr]^{(0)}
\hspace{2.cm}\\
 = \; \left( -1 \right)^{j_m+j_n+k'+x+1} \sqrt{\left[ k, k', x \right]}
\left\{
    \begin{array}{ccc}
      j_m & j_n & k \\
      x   & k'  & j_{n}
    \end{array}
\right\}
\sum_{J} \left( -1 \right)^{J} \sqrt{\left[ J \right]}
\left\{
    \begin{array}{ccc}
      j_m & j_m & J \\
      j_n & j_n & k'
    \end{array}
\right\}
\\
\times
 \; \Bigl[ \bigl[ a^{(j_m)}_1 \times \tilde{a}^{(j_m)}_6 \bigr]^{(J)} \times \bigl[ \tilde{a}^{(j_n)}_3 \times  a^{(j_n)}_5  \bigr]^{(J)} \Bigr]^{(0)} 
\\
+ \; 
\left( -1 \right)^{j_m+j_n+k'} \sqrt{\left[ k, k' \right]} \sum_{J_1,J_2} \sqrt{\left[ J_1, J_2 \right]}
\left\{
    \begin{array}{ccc}
      j_m & j_n & k \\
			J_1 & J_2 & x \\
      j_m & j_n & k'
    \end{array}
\right\}
\\
\times
\biggl[ \bigl[ a^{(j_m)}_1 \times \tilde{a}^{(j_m)}_6 \bigr]^{(J_1)} \times \Bigl[ \bigl[ a^{(j_n)}_4 \times \tilde{a}^{(j_n)}_3 \bigr]^{(x)} \times \bigl[ \tilde{a}^{(j_n)}_2 \times a^{(j_n)}_5 \bigr]^{(J_2)}  \Bigr]^{(J_1)} \biggr]^{(0)}
 \\
 + \; \sqrt{\left[ j_n, k \right]} \sum_{J} \left( -1 \right)^{J} \sqrt{\left[ J \right]}
\left\{
    \begin{array}{ccc}
      j_m & j_m & J \\
      j_n & j_n & k
    \end{array}
\right\}
\\
\times
 \Bigl[ \bigl[ a^{(j_m)}_1 \times \tilde{a}^{(j_m)}_6 \bigr]^{(J)} \times \bigl[  \tilde{a}^{(j_n)}_2 \times a^{(j_n)}_5\bigr]^{(J)} \Bigr]^{(0)} \delta \left( x,0 \right) \; \delta \left( k,k' \right).
\end{split}
\end{equation}
The first and the third terms in the expression (\ref{eq:Tensor17}) correspond to the spin-angular part of the two-particle operators. The program library \texttt{librang}~\cite{Gaigalas:2022} is therefore sufficient for their calculation. The second term, with its three pairs of second quantization operators, corresponds to the three-particle operator (\ref{eq:Tensor12}). In this case, we have succeeded in obtaining a tensorial expression for this three-particle operator for which the library \texttt{librang}~\cite{Gaigalas:2022} extension given in the paper~\cite{Gaigetal:2025VVT} is sufficient (see section 3.1 of the paper~\cite{Gaigetal:2025VVT} for details). Only instead of formula (12) from paper~\cite{Gaigetal:2025VVT} we need to use the expression:

\begin{eqnarray}
\label{eq:tg}
\hspace{-2.0cm}
   \redmem{(n_n\ell_n)\, j_n^w\, \alpha J}
	        {\, \Bigl[ \bigl[ a^{(j_n)} \times \tilde{a}^{(j_n)} \bigr]^{(x)}  \times \bigl[ \tilde{a}^{(j_n)} \times a^{(j_n)} \bigr]^{(J_2)} \Bigr]^{(J_1)} \,}
					{(n_n\ell_n)\, j_n^{w} \, \alpha ^{\prime }J^{\prime }}
   \nonumber  \\[1ex]
\hspace{-1.5cm}
   =\left( -1\right) ^{J + J^{\prime } + J_1} \ \sqrt{\left[ J_1 \right]} \ 
   \displaystyle {\sum_{\alpha ^{\prime \prime }J^{\prime \prime }}}
   \ \left\{
   \begin{array}{ccc}
      x         & J_2 & J_1 \\
      J^{\prime } & J   & J^{\prime \prime }
   \end{array}
   \right\}
   \nonumber  \\[1ex]
\hspace{-1cm}
   \times
   \redmem{(n_n\ell_n)\, j_n^w \,\alpha J}{\, \bigl[ a^{(j_n)} \times \tilde{a}^{(j_n)} \bigr]^{(x)} \,}
	        {(n_n\ell_n)\, j_n^{w}\, \alpha ^{\prime \prime }J^{\prime \prime }} \
   \nonumber  \\[1ex]
\hspace{-1cm}
   \times
   \redmem{(n_n\ell_n)\, j_n^{w}\, \alpha ^{\prime \prime }J^{\prime \prime }}
	        {\, \bigl[ \tilde{a}^{(j_n)} \times a^{(j_n)} \bigr]^{(J_2)} \,}{(n_n\ell_n)\, j_n^{w}\, \alpha ^{\prime }J^{\prime }}.
\end{eqnarray}
Therefore, the library \texttt{librang} must be extended by programming the expression (\ref{eq:tg}).

The above describes how to calculate the reduced matrix elements of a three-particle Feynman diagram CV$_7$ in the general case for the third type of core-valence correlations. But it is possible to extract the individual parts which are more straightforward to calculate, i.e. the coefficient $\Delta \mathcal{E}_0$ (does not depend on the term), $\Delta \mathcal{F}^{k}(n,n)$, and $\Delta \mathcal{F}^{k}(m,n)$ (regular spin-angular library \texttt{librang} can be used). The extraction of these expressions is the same as in papers~\cite{Gaigetal:2024CV,Gaigetal:2024C,Gaigetal:2024CC,Gaigetal:2025VV,Gaigetal:2025VVT}. Therefore, only the reduced matrix element of part of Feynman diagrams $A_5$ with tensorial product (\ref{eq:Tensor12}) is calculated from the general expression when the rank $k > 0$. The part which is calculated from the general expression will be denoted by $\Delta \widetilde{\widetilde{\mathcal{R}}}^{(k, k', x)} \left( m n n  \right)$ in the future.

\subsection{The spin-angular part of the fourth type of core-valence correlations}
\label{sec:PT_SA_Fourth}

Two different Feynman diagrams $A_5$ and $A_6$ from Fig.~\ref{CV_4_type} describe this type of correlation. We will consider each separately, as their spin-angular part differs significantly.

\subsubsection{The direct part of the fourth type of core-valence correlations}
\label{sec:PT_SA_direct_Fourth}

Now let us discuss the diagram $A_5$, with tensorial structure
\begin{equation}
\label{eq:Tensor211}
\biggl[ \Bigl[ \bigl[ a^{(j_m)}_1 \times \tilde{a}^{(j_n)}_2 \bigr]^{(k)}   \times \bigl[ \tilde{a}^{(j_p)}_3 \times a^{(j_p)}_4 \bigr]^{(x)} \Bigr]^{(k')}  \times \bigl[ a^{(j_n)}_5 \times \tilde{a}^{(j_m)}_6 \bigr]^{(k')} \biggr]^{(0)}.
\end{equation}
As in (\ref{eq:Tensor11}), (\ref{eq:Tensor211}) is unsuitable to calculate the spin-angular part because there are two pairs of secondary quantization operators acting on different subshells, i.e., $\bigl[ a^{(j_m)}_1 \times \tilde{a}^{(j_n)}_2 \bigr]^{(k)}$ and $\bigl[ a^{(j_n)}_5 \times \tilde{a}^{(j_m)}_6 \bigr]^{(k)}$. Therefore, this tensorial product has to be transformed into the following, using the commutation rules of secondary quantization
\begin{equation}
\label{eq:Tensor212}
\biggl[ \Bigl[ \bigl[ a^{(j_m)}_1 \times \tilde{a}^{(j_m)}_6 \bigr]^{(J_1)} \times  \bigl[ \tilde{a}^{(j_p)}_3 \times a^{(j_p)}_4 \bigr]^{(x)} \Bigr]^{(J_2)} \times \bigl[ \tilde{a}^{(j_n)}_2 \times a^{(j_n)}_5 \bigr]^{(J_2)}  \biggr]^{(0)}.
\end{equation}

But first of all, using the commutation rule (\ref{eq:Tensor13}) of secondary quantization we interchange the operators $a_1$ and $a_2$ and the operators $a_5$ and $a_6$. This leads to the expression
\begin{equation}
\label{eq:Tensor213}
\begin{split}
\biggl[ \Bigl[ \bigl[ a^{(j_m)}_1 \times \tilde{a}^{(j_n)}_2 \bigr]^{(k)}   \times \bigl[ \tilde{a}^{(j_p)}_3 \times a^{(j_p)}_4 \bigr]^{(x)} \Bigr]^{(k')}  \times \bigl[ a^{(j_n)}_5 \times \tilde{a}^{(j_m)}_6 \bigr]^{(k')} \biggr]^{(0)}
\\
= \; \left( -1 \right)^{k+k'}
\biggl[ \Bigl[ \bigl[ \tilde{a}^{(j_n)}_2 \times a^{(j_m)}_1 \bigr]^{(k)}   \times \bigl[ \tilde{a}^{(j_p)}_3 \times a^{(j_p)}_4 \bigr]^{(x)} \Bigr]^{(k')}  \times \bigl[ \tilde{a}^{(j_m)}_6 \times a^{(j_n)}_5 \bigr]^{(k')} \biggr]^{(0)} .
\end{split}
\end{equation}

Now transforming the tensorial product (\ref{eq:Tensor213}) to (\ref{eq:Tensor212}) is determined by the tensorial product recoupling matrix, which is represented by the diagram $B_3$ from~\cite{Gaigetal:2025VVT} (see Fig. 5 of the paper~\cite{Gaigetal:2025VVT}). The final result is:

\begin{equation}
\label{eq:Tensor214}
\begin{aligned}
\biggl[ \Bigl[ \bigl[ a^{(j_m)}_1 \times \tilde{a}^{(j_n)}_2 \bigr]^{(k)}   \times \bigl[ \tilde{a}^{(j_p)}_3 \times a^{(j_p)}_4 \bigr]^{(x)} \Bigr]^{(k')}  \times \bigl[ a^{(j_n)}_5 \times \tilde{a}^{(j_m)}_6 \bigr]^{(k')} \biggr]^{(0)}
\\
= \; \left( -1 \right)^{k+k'} \sum_{J_1,J_2}  \; B_3 \;
\biggl[ \Bigl[ \bigl[ a^{(j_m)}_1 \times \tilde{a}^{(j_m)}_6 \bigr]^{(J_1)} \times  \bigl[ \tilde{a}^{(j_p)}_3 \times a^{(j_p)}_4 \bigr]^{(x)} \Bigr]^{(J_2)} \times \bigl[ \tilde{a}^{(j_n)}_2 \times a^{(j_n)}_5 \bigr]^{(J_2)}  \biggr]^{(0)}
\\
= \;
\left( -1 \right)^{j_m+j_n+k'+x} \sqrt{\left[ k, k' \right]} \sum_{J_1,J_2} \sqrt{\left[ J_1, J_2 \right]}
\left\{
    \begin{array}{ccc}
      j_m & j_n & k \\
			J_1 & J_2 & x \\
      j_m & j_n & k'
    \end{array}
\right\}
\\
\times \; \biggl[ \Bigl[ \bigl[ a^{(j_m)}_1 \times \tilde{a}^{(j_m)}_6 \bigr]^{(J_1)} \times  \bigl[ \tilde{a}^{(j_p)}_3 \times a^{(j_p)}_4 \bigr]^{(x)} \Bigr]^{(J_2)} \times \bigl[ \tilde{a}^{(j_n)}_2 \times a^{(j_n)}_5 \bigr]^{(J_2)}  \biggr]^{(0)}.
\end{aligned}
\end{equation}

The program library \texttt{librang}~\cite{Gaigalas:2022} available in the {\sc Grasp} package is sufficient for the calculation of the reduced matrix element of Feynman diagrams $A_5$, but the method of calculation of this type of operator is not described in the paper \cite{Gaigalas:2022}.
This is described in the paper~\cite{Gaigetal:2025VVT} (see subsection~3.2.1 of~\cite{Gaigetal:2025VVT}).

\subsubsection{The exchange part of the fourth type of core-valence correlations}
\label{sec:PT_SA_exchange_Fourth}

Now let us discuss the diagram $A_6$, with tensorial structure
\begin{equation}
\label{eq:Tensor311}
\biggl[ \Bigl[ \bigl[ a^{(j_m)}_1 \times \tilde{a}^{(j_n)}_2 \bigr]^{(k)}   \times \bigl[ \tilde{a}^{(j_p)}_3 \times a^{(j_n)}_4 \bigr]^{(x)} \Bigr]^{(k')}  \times \bigl[ a^{(j_p)}_5 \times \tilde{a}^{(j_m)}_6 \bigr]^{(k')} \biggr]^{(0)}.
\end{equation}
This tensorial product is unsuitable to calculate the spin-angular part for the same reason as for (\ref{eq:Tensor11}) and (\ref{eq:Tensor211}). Therefore, this tensorial product has to be transformed into the following, using the commutation rules of secondary quantization
\begin{equation}
\label{eq:Tensor312}
\biggl[ \Bigl[ \bigl[ a^{(j_m)}_1 \times \tilde{a}^{(j_m)}_6 \bigr]^{(J_1)} \times  \bigl[ \tilde{a}^{(j_p)}_3 \times a^{(j_p)}_5 \bigr]^{(y)} \Bigr]^{(J_2)} \times \bigl[ \tilde{a}^{(j_n)}_2 \times a^{(j_n)}_4 \bigr]^{(J_2)}  \biggr]^{(0)}.
\end{equation}
But first of all, using the commutation rule (\ref{eq:Tensor13}) of secondary quantization we interchange the operators $a_1$ and $a_2$ and the operators $a_5$ and $a_6$. This leads to the expression
\begin{equation}
\label{eq:Tensor313}
\begin{aligned}
\biggl[ \Bigl[ \bigl[ a^{(j_m)}_1 \times \tilde{a}^{(j_n)}_2 \bigr]^{(k)}   \times \bigl[ \tilde{a}^{(j_p)}_3 \times a^{(j_n)}_4 \bigr]^{(x)} \Bigr]^{(k')}  \times \bigl[ a^{(j_p)}_5 \times \tilde{a}^{(j_m)}_6 \bigr]^{(k')} \biggr]^{(0)}
\\
= \; \left( -1 \right)^{j_n + j_p + k + k' +1}
\biggl[ \Bigl[ \bigl[ \tilde{a}^{(j_n)}_2 \times a^{(j_m)}_1 \bigr]^{(k)}   \times \bigl[ \tilde{a}^{(j_p)}_3 \times a^{(j_n)}_4 \bigr]^{(x)} \Bigr]^{(k')}  \times \bigl[ \tilde{a}^{(j_m)}_6 \times a^{(j_p)}_5 \bigr]^{(k')} \biggr]^{(0)} .
\end{aligned}
\end{equation}

Now transforming the tensorial product (\ref{eq:Tensor313}) to (\ref{eq:Tensor312}) is determined by the tensorial product recoupling matrix, which is represented by the diagram $B_4$ from~\cite{Gaigetal:2025VVT} (see Fig. 6 of the paper~\cite{Gaigetal:2025VVT}). The final result is:

\begin{equation}
\label{eq:Tensor314}
\begin{aligned}
\biggl[ \Bigl[ \bigl[ a^{(j_m)}_1 \times \tilde{a}^{(j_n)}_2 \bigr]^{(k)}   \times \bigl[ \tilde{a}^{(j_p)}_3 \times a^{(j_n)}_4 \bigr]^{(x)} \Bigr]^{(k')}  \times \bigl[ a^{(j_p)}_5 \times \tilde{a}^{(j_m)}_6 \bigr]^{(k')} \biggr]^{(0)}
\\
= \; \left( -1 \right)^{j_n + j_p + k + k' +1} \sum_{J_1,J_2,y} B_4 \; 
\biggl[ \Bigl[ \bigl[ a^{(j_m)}_1 \times \tilde{a}^{(j_m)}_6 \bigr]^{(J_1)} \times  \bigl[ \tilde{a}^{(j_p)}_3 \times a^{(j_p)}_5 \bigr]^{(y)} \Bigr]^{(J_2)} \times \bigl[ \tilde{a}^{(j_n)}_2 \times a^{(j_n)}_4 \bigr]^{(J_2)}  \biggr]^{(0)}
\\
= \; \left( -1 \right)^{j_m + j_n  + k +1} \sqrt{\left[ k, k', x \right]} \sum_{J_1,J_2,y} \left( -1 \right)^{J_1+J_2} \sqrt{\left[ J_1, J_2, y \right]}
\\
\times \; 
\biggl[ \Bigl[ \bigl[ a^{(j_m)}_1 \times \tilde{a}^{(j_m)}_6 \bigr]^{(J_1)} \times  \bigl[ \tilde{a}^{(j_p)}_3 \times a^{(j_p)}_5 \bigr]^{(y)} \Bigr]^{(J_2)} \times \bigl[ \tilde{a}^{(j_n)}_2 \times a^{(j_n)}_4 \bigr]^{(J_2)}  \biggr]^{(0)}
\\
\times \; \sum_{z} \left[ z \right]
		  \left\{
    \begin{array}{ccc}
      J_{1} & j_{n} & z \\
      j_{n} & y     & J_{2}
    \end{array} \right\}
		  \left\{
    \begin{array}{ccc}
      y & j_{n} & z \\
      x & j_{p} & j_{p}
    \end{array} \right\}
		  \left\{
		 \begin{array}{ccc}
      j_{p} & x     & z \\
      k     & j_{m} & k'
    \end{array} \right\}
		  \left\{
		 \begin{array}{ccc}
      j_{m} & k     & z \\
      j_{n} & J_{1} & j_{m}
    \end{array} \right\}.
\end{aligned}
\end{equation}

The program library \texttt{librang}~\cite{Gaigalas:2022} available in the {\sc Grasp} package is sufficient for the calculation of the reduced matrix element of Feynman diagram $A_6$. The use of the library in this case is exactly the same as the one presented in~\cite{Gaigetal:2025VVT} (see subsection~3.2.2 of~\cite{Gaigetal:2025VVT}). In this case only the multiplier 
$\Theta ^{\prime }\left( n_i\lambda _i,n_j\lambda _j,n_i^{\prime
}\lambda _i^{\prime },n_j^{\prime }\lambda _j^{\prime },\Xi \right) $ differs.

\subsubsection{The special cases of the fourth type of core-valence correlations}
\label{sec:PT_SA_special_Fourth}

The way to calculate reduced matrix elements of Feynman diagrams $A_5$ and $A_6$ in the general case is presented in subsections~\ref{sec:PT_SA_direct_Fourth}, and \ref{sec:PT_SA_exchange_Fourth}. But it is possible to extract, as it was shown in subsection~\ref{sec:PT_SA_Third}, the individual parts which are more straightforward to calculate, i.e., the coefficient $\Delta \mathcal{E}_0$ (does not depend on the term) and $\Delta \mathcal{F}^{k}(n,p)$ (regular spin-angular library \texttt{librang} can be used). The extraction of these expressions is the same as in the papers~\cite{Gaigetal:2024CV,Gaigetal:2024C,Gaigetal:2024CC,Gaigetal:2025VV,Gaigetal:2025VVT}. Therefore, only the reduced matrix element of part of three-particle Feynman diagrams $A_5$ and $A_6$ is calculated from the general expression, where the rank $k > 0$. The part which is calculated from the general expression will be denoted by $\Delta \widetilde{\widetilde{\mathcal{R}}}^{(k, k', x)} \left( m n p  \right)$ in the future.

In the next section, we will give the final expressions for these two types of correlations (third and fourth types of core-valence correlations), as we did in the papers~\cite{Gaigetal:2024CV,Gaigetal:2024C,Gaigetal:2024CC,Gaigetal:2025VV,Gaigetal:2025VVT}.

\section{Combination of RCI approximation
with the stationary second-order Rayleigh-Schr\"odinger many-body perturbation theory}
\label{Sec:VVImplementation}

Similar to core-valence~\cite{Gaigetal:2024CV}, core~\cite{Gaigetal:2024C}, core-core~\cite{Gaigetal:2024CC}, and valence-valence~\cite{Gaigetal:2025VV,Gaigetal:2025VVT} correlations the admixed configurations from core-valence correlations deriving from three-particle Feynman diagram CV$_7$
(Eqs. (\ref{eq:CVT-a}) and (\ref{eq:CVT-b}))
can be added to usual energy $E_0 \left(K \right)$ of the 
term $\chi J$ of the configuration $K$ and can be
expressed as the energy $\Delta \mathcal{E}_0 \left(K J \right)$, which does not depend on the term, and the sum of the product of Slater integrals and spin-angular coefficients, describing the interaction within and between open subshells:
\begin{eqnarray}
\label{eq:CVBogEnergy}
\hspace*{-2.5cm}
   E\left(K \chi J \right)
	\nonumber \\
& &
   = E_0 \left(K J\right) + \Delta \mathcal{E}_0 \left(K J \right) 
	\nonumber \\  [0.2cm]
& &
	+ \; \sum_{n\ell j} \sum_{k>0} \widetilde{f}_k \left( \ell j^{w}, \; K \chi J  \right)
	\left[ \mathcal{F}^{k} \left( n \ell j, \; n \ell j \right)  
	+ \Delta \mathcal{F}^{k} \left( n \ell j, \; n \ell j \right) \right]
	\nonumber \\
& &
	+ \; \sum_{n\ell j} \sum_{n'\ell'j' > n\ell j} \left\{ \sum_{k>0} \widetilde{f}_k \left( \ell j^{w} \; \ell' j'^{w'},
	\; K \chi J  \right) \right.
%  \nonumber \\
%  \times 
\left[ \mathcal{F}^{k} \left( n \ell j, \; n '\ell' j' \right)  
	+ \Delta \mathcal{F}^{k} \left( n \ell j, \; n' \ell' j' \right) \right]
	\nonumber \\ [0.2cm]
& & 
	+ \sum_{k} \widetilde{g}_k \left( \ell j^{w} \; \ell' j'^{w'}, \; K \chi J  \right)	
%	\nonumber \\
%  \times 
\mathcal{G}^{k} \left( n \ell j, \; n '\ell' j' \right)  
	\nonumber \\ [0.2cm]
& &
\left.
	+ \sum_{k} \widetilde{v}_k \left( \ell j^{w} \; \ell' j'^{w'}, \ell j^{w-2} \; \ell' j'^{w'+2},
	\; K \chi J \; K' \chi' J \right)	
%	\nonumber \\
%& &
%  \times 
%\left[
\mathcal{R}^{k} \left( n \ell j n \ell j, \; n '\ell' j' n '\ell' j' \right) \right\} 
%  \right.
	\nonumber \\
& & 
%+ \sum_{n\ell j} \; \sum_{n'\ell'j' \neq n\ell j} \; \sum_{n''\ell''j'' \neq n\ell j} 
+ \sum_{\substack{n\ell j \\ n'\ell'j' \, \neq \, n\ell j}}
\; \sum_{\substack{k>0 \\ k',x}} 
\left< \Psi \left\| \biggl[ \bigl[ a^{(j)} \times \tilde{a}^{(j)} \bigr]^{(k)} \times  \Bigl[ \bigl[ a^{(j')} \times \tilde{a}^{(j')} \bigr]^{(x)} \times \bigl[ \tilde{a}^{(j')} \times a^{(j')} \bigr]^{(k')} \Bigr]^{(k)} \biggr]^{(0)} \right\| \Psi \right> 
	\nonumber \\
& & 
\hspace{1.0cm} \times \;
\Delta \widetilde{\widetilde{\mathcal{R}}}^{(k,k',x)}
\left( n \ell j \; n '\ell' j' \; n '\ell' j' \right)
	\nonumber \\
& & 
%+ \sum_{n\ell j} \; \sum_{n'\ell'j' \neq n\ell j} \; \sum_{n''\ell''j'' \neq n\ell j} 
+ \sum_{\substack{n\ell j \\ n'\ell'j' \, \neq \, n\ell j \\ n''\ell''j'' \, \neq \, n\ell j}}
\; \sum_{\substack{k>0 \\ k',x}} 
\left< \Psi \left\| \biggl[ \Bigl[ \bigl[ a^{(j)} \times \tilde{a}^{(j)} \bigr]^{(k)} \times \bigl[ \tilde{a}^{(j'')} \times a^{(j'')} \bigr]^{(x)} \Bigr]^{(k')} \times \bigl[ \tilde{a}^{(j')} \times a^{(j')} \bigr]^{(k')} \biggr]^{(0)} \right\| \Psi \right> 
	\nonumber \\
& & 
\hspace{1.0cm} \times \;
\Delta \widetilde{\widetilde{\mathcal{R}}}^{(k,k',x)}
\left( n \ell j \; n '\ell' j' \; n ''\ell'' j'' \right),
\end{eqnarray}
where $\widetilde{f}_k$, $\widetilde{g}_k$, and $\widetilde{v}_k$ are spin-angular coefficients from which submatrix elements $\redmem{\ell j}{\, C^{(k)} \,}{ \ell^{\prime} j^{\prime}}$ are extracted. 
Therefore, summation over $k$ runs over all 
possible values instead of the values that satisfy the triangular condition $\left( \ell \ell^{\prime} k\right)$ as it is in the regular case. The $\mathcal{F}^{k} \left( n \ell j, \; n '\ell' j' \right)$,
$\mathcal{G}^{k} \left( n \ell j, \; n '\ell' j' \right)$, and $\mathcal{R}^{k} \left( n \ell j n \ell j, \; n '\ell' j' n '\ell' j' \right)$ are generalized integrals of electrostatic interaction between electrons. The definition of $\mathcal{R}^{k} \left( n \ell j n \ell j, \; n '\ell' j' n '\ell' j' \right)$ is the following:
\begin{eqnarray}
\label{eq:BogRk}
\hspace*{-2.5cm}
   \mathcal{R}^{k}\left(i j, i' j'\right)
	\nonumber \\
& &
   = \left\{ \left[ 1 + \delta \left( i, j \right) \right]  \left[ 1 + \delta \left( i', j' \right) \right]  \right\} ^{-1/2}
	 \, R^{k}\left(n_i j_i \, n_jj_j, \, n_{i'}j_{i'} \, n_{j'}j_{j'} \right)
\nonumber \\
& &
	   \times \redmem{\ell_i j_{i}}{\, C^{(k)} \,}{ \ell_{i'} j_{i'}}
     \redmem{\ell_j j_{j}}{\, C^{(k)} \,}{ \ell_{j'} j_{j'}}, 
\end{eqnarray}
where $R^{k}\left(n_i j_i \, n_jj_j, \, n_{i'}j_{i'} \, n_{j'}j_{j'} \right)$ is the same radial integral as in 
Eq. (\ref{eq:deffX}). Definitions  $\mathcal{F}^{k} \left( n \ell j, \; n '\ell' j' \right)$,
$\mathcal{G}^{k} \left( n \ell j, \; n '\ell' j' \right)$ straightforwardly follow from Eq. (\ref{eq:BogRk}).
Due to the inherent complexity of the three-particle operator, the expression (\ref{eq:CVBogEnergy}) does not fully distinguish all the members that are
independent of the term. Therefore, a very small number of them are retained in the members 
$\Delta \mathcal{F}^{k} \left( n \ell j, \; n \ell j \right)$,
$\Delta \widetilde{\widetilde{\mathcal{R}}}^{(k,k',x)}\left( n \ell j \; n '\ell' j' \; n '\ell' j' \right)$, and $\Delta \widetilde{\widetilde{\mathcal{R}}}^{(k,k',x)}\left( n \ell j \; n '\ell' j' \; n ''\ell'' j'' \right)$.

The contribution deriving from the core-valence correlations of the configurations $K'$ to $E (K \chi J)$ in the second-order of the perturbation theory
can be extracted from Eq. (\ref{eq:CVBogEnergy}) as
\begin{eqnarray}
\label{eq:BogEnergy_PT}
\hspace*{-2.5cm}
  \Delta E_{PT (CV T)}
	\nonumber \\
& &
   =  \Delta \mathcal{E}_0 \left(K J \right) 
	\nonumber \\  [0.2cm]
& &
	+ \; \sum_{n\ell j} \sum_{k>0} \widetilde{f}_k \left( \ell j^{w}, \; K \chi J  \right)
	\Delta \mathcal{F}^{k} \left( n \ell j, \; n \ell j \right)
	\nonumber \\
& &
	+ \; \sum_{n\ell j} \sum_{n'\ell'j' > n\ell j} \sum_{k>0} \widetilde{f}_k \left( \ell j^{w} \; \ell' j'^{w'},
	\; K \chi J  \right)
%  \nonumber \\
%  \times 
 \Delta \mathcal{F}^{k} \left( n \ell j, \; n' \ell' j' \right) 
	\nonumber \\ [0.2cm]
& &
%+ \sum_{n\ell j} \; \sum_{n'\ell'j' \neq n\ell j} \; \sum_{n''\ell''j'' \neq n\ell j} 
+ \sum_{\substack{n\ell j \\ n'\ell'j' \, \neq \, n\ell j}}
\; \sum_{\substack{k>0 \\ k',x}} 
\left< \Psi \left\| \biggl[ \bigl[ a^{(j)} \times \tilde{a}^{(j)} \bigr]^{(k)} \times  \Bigl[ \bigl[ a^{(j')} \times \tilde{a}^{(j')} \bigr]^{(x)} \times \bigl[ \tilde{a}^{(j')} \times a^{(j')} \bigr]^{(k')} \Bigr]^{(k)} \biggr]^{(0)} \right\| \Psi \right> 
	\nonumber \\
& & 
\hspace{1.0cm} \times \;
\Delta \widetilde{\widetilde{\mathcal{R}}}^{(k,k',x)}
\left( n \ell j \; n '\ell' j' \; n '\ell' j' \right)
	\nonumber \\
& & 
%+ \sum_{n\ell j} \; \sum_{n'\ell'j' \neq n\ell j} \; \sum_{n''\ell''j'' \neq n\ell j} 
+ \sum_{\substack{n\ell j \\ n'\ell'j' \, \neq \, n\ell j \\ n''\ell''j'' \, \neq \, n\ell j}}
\; \sum_{\substack{k>0 \\ k',x}} 
\left< \Psi \left\| \biggl[ \Bigl[ \bigl[ a^{(j)} \times \tilde{a}^{(j)} \bigr]^{(k)} \times \bigl[ \tilde{a}^{(j'')} \times a^{(j'')} \bigr]^{(x)} \Bigr]^{(k')} \times \bigl[ \tilde{a}^{(j')} \times a^{(j')} \bigr]^{(k')} \biggr]^{(0)} \right\| \Psi \right> 
	\nonumber \\
& & 
\hspace{1.0cm} \times \;
\Delta \widetilde{\widetilde{\mathcal{R}}}^{(k,k',x)}
\left( n \ell j \; n '\ell' j' \; n ''\ell'' j'' \right).
\end{eqnarray}

\begin{table*}
\begin{center}
%\begin{tiny}
%begin{small}
\begin{tabular}{|l|} \hline
$\Delta \mathcal{E}_0$ corrections \\ \hline \hline
\\
% First type
$\overbrace{(n_{a} \ell_{a})\, j_{a}^{2j_a+1}}^{\text{core subshell}} \; \overbrace{(n_{m} \ell_{m})\, j_{m}^{w_m} \, (n_{n} \ell_{n})\, j_{n}^{w_n}}^{\text{valence subshells}} \;
\rightarrow \; \overbrace{(n_{a} \ell_{a})\, j_{a}^{2j_a}}^{\text{core subshell}} \; \overbrace{(n_{m} \ell_{m})\, j_{m}^{w_m-1} \, (n_{n} \ell_{n})\, j_{n}^{w_n+2}}^{\text{valence subshells}}$ \\
\\
$\underbrace{2 \; \left( -1 \right)^{j_m+j_a} \; \frac{ w_m \left( \left[ j_n \right] - 2w_n \right)}{\left[ j_m, j_n\right]} \; \sum_{k} \;
\Bigl\{ \mathcal{C}\left( k, \, n n, \, a m \right) + \frac{1}{\left[ k \right]} \, \mathcal{P}\left( kk, \, m a, \, n n \right) \Bigr\}}_{\text{from $\text{CV}_7$ Feynman diagram}}$ \\
\\ \hline \hline
\\
% Second type
$\overbrace{(n_{a} \ell_{a})\, j_{a}^{2j_a+1}}^{\text{core subshell}} \; \overbrace{(n_{m} \ell_{m})\, j_{m}^{w_m} \; (n_{n} \ell_{n})\, j_{n}^{w_n} \; (n_{p} \ell_{p})\, j_{p}^{w_p}}^{\text{valence subshells}} \;
\rightarrow \; \overbrace{(n_{a} \ell_{a})\, j_{a}^{2j_a}}^{\text{core subshell}} \; \overbrace{(n_{m} \ell_{m})\, j_{m}^{w_m-1} \, (n_{n} \ell_{n})\,j_{n}^{w_n+1}  \; (n_{p} \ell_{p})\, j_{p}^{w_p+1}}^{\text{valence subshells}}$ \\
\\ $\underbrace{- \left( -1 \right)^{j_m+j_n+j_p+j_a}\frac{w_m \left( \left[ j_n \right] - w_n \right) \left( \left[ j_p \right] - w_p \right)}{\left[ j_m, j_n, j_p \right]} \; \sum_{k}  \left\{ \frac{ \mathcal{P}\left( kk, \, m a, \, n p \right)}{\left[ k\right]} + \mathcal{C}\left( k, \, m a, \, n p \right) \right\} \biggl( 1 + \mbox{P}\left( n \rightleftharpoons p \right) \biggr)
	}_{\text{from $\text{CV}_7$ Feynman diagram}}$ \\
\\ \hline 
\end{tabular}
\end{center}
\caption{Expressions for core-valence corrections to the energy in Eqs. (\ref{eq:CVBogEnergy}) and (\ref{eq:BogEnergy_PT}), independent of the term.}
\label{tab:Implemen_CV1}
%\end{small}
\end{table*}

\begin{table*}
\begin{center}
%\begin{tiny}
%begin{small}
\begin{tabular}{|lcc|} \hline
Corrections & Slater integral & $k$ values\\ \hline  \hline
& & \\
% First type
\multicolumn{3}{|c|}{$\overbrace{(n_{a} \ell_{a})\, j_{a}^{2j_a+1}}^{\text{core subshell}} \; \overbrace{(n_{m} \ell_{m})\, j_{m}^{w_m} \, (n_{n} \ell_{n})\, j_{n}^{w_n}}^{\text{valence subshells}} \;
\rightarrow \; \overbrace{(n_{a} \ell_{a})\, j_{a}^{2j_a}}^{\text{core subshell}} \; \overbrace{(n_{m} \ell_{m})\, j_{m}^{w_m-1} \, (n_{n} \ell_{n})\, j_{n}^{w_n+2}}^{\text{valence subshells}}$} \\
& &\\
$\underbrace{ 
-4 \; \frac{\left[ k \right]}{\left[ j_m \right]} \; w_m \; \mathcal{A^{\prime}}\left( k, \, n n, \, m a \right) }_{\text{from $\text{CV}_{7}$ Feynman diagram}}$
&$\Delta \mathcal{F}^{k}(n,n)$ & $k \geq 0$\\
& &\\
& &\\
$\underbrace{ 
2 \left[ k \right] \sum_{k'} \left( -1 \right)^{j_n+j_a+k'} 
		  \left\{
    \begin{array}{ccc}
      j_{m} & j_{m} & k \\
      j_{n} & j_{n} & k'
    \end{array} \right\}
\; \times}_{\text{from $\text{CV}_{7}$ Feynman diagram}}$
&$\Delta \mathcal{F}^{k}(m,n)$ & $k>0$ \\
%& & \\
& & \\
\hspace{0.5cm} $\underbrace{\times \;  \left\{
		\mathcal{C}\left( k', \, n n, \, a m \right)
		+ \frac{1}{\left[ k' \right]} \, 
%		  \left\{
%    \begin{array}{ccc}
%      j_{m} & j_{m} & k \\
%      j_{n} & j_{n} & k'
%    \end{array} \right\}  			
		\mathcal{P}\left( k'k', \, m a, \, n n \right)
    \right\}		
}_{\text{from $\text{CV}_{7}$ Feynman diagram}}$ 
& & \\
& &\\
& &\\
%\multicolumn{3}{|l|}{
$\underbrace{ 
2 
%\sum_{k,k', x} 
\left( -1 \right)^{j_m + j_n} \sqrt{\left[ k, k', x \right]} \; \mathcal{G^{\prime}}\left( k \, k' \, x, \, n n, \, m a \right)}_{\text{from $\text{CV}_{7}$ Feynman diagram}}$
&$\Delta \widetilde{\widetilde{\mathcal{R}}}^{(k, k', x)} \left( m n n \right)$ & $k>0$ \\
& & \\ \hline 
\end{tabular}
\end{center}
\caption{Expressions for Slater integrals $\Delta \mathcal{F}^{k}(n,n)$, $\Delta \mathcal{F}^{k}(m,n)$, and $\Delta \widetilde{\widetilde{\mathcal{R}}}^{(k, k', x)} \left( m n n  \right)$
(see Eqs. (\ref{eq:CVBogEnergy}) and (\ref{eq:BogEnergy_PT})) corrections corresponding 
to the third type of core-valence correlations.} 
\label{tab:Implemen_CV2}
%\end{small}
\end{table*}

\begin{table*}
\begin{center}
%\begin{tiny}
%begin{small}
\begin{tabular}{|lcc|} \hline
Corrections & Slater integral & $k$ values\\ \hline  \hline
& & \\
% Second type
\multicolumn{3}{|c|}{$\overbrace{(n_{a} \ell_{a})\, j_{a}^{2j_a+1}}^{\text{core subshell}} \; \overbrace{(n_{m} \ell_{m})\, j_{m}^{w_m} \; (n_{n} \ell_{n})\, j_{n}^{w_n} \; (n_{p} \ell_{p})\, j_{p}^{w_p}}^{\text{valence subshells}} \;
\rightarrow \; \overbrace{(n_{a} \ell_{a})\, j_{a}^{2j_a}}^{\text{core subshell}} \; \overbrace{(n_{m} \ell_{m})\, j_{m}^{w_m-1} \, (n_{n} \ell_{n})\,j_{n}^{w_n+1}  \; (n_{p} \ell_{p})\, j_{p}^{w_p+1}}^{\text{valence subshells}}$} \\
& &\\
$\underbrace{ 
 \left( -1 \right)^{j_m+j_n} \, \left[k\right]\sqrt{\left[ k \right]} \, \frac{w_m}{\sqrt{\left[j_m\right]}} \; \Biggl\{ \left( -1 \right)^{k+1}  \; \mathcal{G^{\prime}}\left( 0 \, k \, k, \, m a, \, n p \right) \Biggr.}_{\text{from $\text{CV}_{7}$ Feynman diagram}}$
&$\Delta \mathcal{F}^{k}(n,p)$ & $k>0$\\
%& &\\
& &\\
$\underbrace{ \Biggl.+ 
 \sum_{k_1, k_2, k_3}
 \left( -1 \right) ^{k_1+k} \left[ k_3 \right] \;
		  \left\{
    \begin{array}{ccc}
      j_{n} & j_{p} & k_3 \\
      k_{1} & k_{2} & j_a
    \end{array} \right\}
\mathcal{Q} \left( k_1k_2, m a, n p \right) }_{\text{from $\text{CV}_{7}$ Feynman diagram} }$ & & \\
%& & \\
& & \\
\hspace{0.5cm} $\underbrace{ \Biggl. \times \;
\mathcal{C}_{12j}\left( j_m j_n j_p, \, k_1 \, k_2 \, k_3, \, 0 \, k \, k \right) \Biggr\} \;
\Biggl( 1 + \mbox{P} 
		  \left( \hspace{-0.15cm}
\begin{array}{lcl}
      n    &\hspace{-0.25cm}\rightleftharpoons&\hspace{-0.25cm}p \\
      k_{1}&\hspace{-0.25cm}\rightleftharpoons&\hspace{-0.25cm}k_{2}
    \end{array}  \hspace{-0.15cm} \right)
%\left( n \rightleftharpoons p \right) \biggr)
 \Biggr) }_{\text{from $\text{CV}_{7}$ Feynman diagram} }$
& &\\
& &\\
& &\\
%\multicolumn{3}{|l|}{
$\underbrace{ 
- 
%\sum_{k,k^{'},x} 
\left( -1 \right)^{j_m + j_n} \sqrt{\left[ k, k^{'}, x \right]} \; \Biggl\{ \left( -1 \right)^{x+1}
\mathcal{G^{\prime}}\left( k \, k^{'} \, x, \, m a, \, n p \right) \Biggr.}_{\text{from $\text{CV}_{7}$ Feynman diagram}}$
&$\Delta \widetilde{\widetilde{\mathcal{R}}}^{(k, k^{'}, x)} \left( m n p \right)$ & $k>0$ \\
& & \\
\hspace{0.5cm} $\underbrace{ 
+ \sum_{k_1, k_2, k_3} \left( -1 \right)^{k + k^{'} + k_1} \left[ k_3 \right]
		  \left\{
    \begin{array}{ccc}
      j_{n} & j_{p} & k_3 \\
      k_{1} & k_{2} & j_a
    \end{array} \right\}
 \mathcal{Q} \left( k_1k_2, m a, n p \right)
		}_{\text{from $\text{CV}_{7}$ Feynman diagram} }$ & & \\
& &\\
%\multicolumn{3}{|l|}{
\hspace{0.5cm} $\underbrace{ \times 
\; \mathcal{C}_{12j}\left( j_m j_n j_p, \, k_1 \, k_2 \, k_3, \, k \, k^{'} \, x \right)\Biggl\}  \;
\Biggl( 1 + \mbox{P} 
		  \left( \hspace{-0.15cm}
\begin{array}{lcl}
      n    &\hspace{-0.25cm}\rightleftharpoons&\hspace{-0.25cm}p \\
      k_{1}&\hspace{-0.25cm}\rightleftharpoons&\hspace{-0.25cm}k_{2} \\
			k' &\hspace{-0.25cm}\rightleftharpoons&\hspace{-0.25cm}x 
    \end{array}  \hspace{-0.15cm} \right)
%\left( n \rightleftharpoons p \right) \biggr)
 \Biggr)}_{\text{from $\text{CV}_{7}$ Feynman diagram}}$ & & \\
& &\\ \hline 
\end{tabular}
\end{center}
\caption{Expressions for Slater integral $\Delta \mathcal{F}^{k}(n,p)$ and $\Delta \widetilde{\widetilde{\mathcal{R}}}^{(k, k', x)} \left( m n p  \right)$ 
(see Eqs. (\ref{eq:CVBogEnergy}) and (\ref{eq:BogEnergy_PT})) corrections corresponding 
to the fourth type of core-valence correlations.} 
\label{tab:Implemen_CV3}
%\end{small}
\end{table*}

The contribution of the third and fourth type of the core-valence correlations in the second-order of the perturbation theory is expressed over $\Delta \mathcal{E}_0 \left(K J \right)$, $\Delta \mathcal{F}^{k} \left( n \ell j, \; n \ell j \right)$, $\Delta \mathcal{F}^{k} \left( n \ell j, \; n' \ell' j' \right)$, $\Delta \widetilde{\widetilde{\mathcal{R}}}^{(k,k',x)}
\left( n \ell j \; n '\ell' j' \; n '\ell' j' \right)$, and $\Delta \widetilde{\widetilde{\mathcal{R}}}^{(k,k',x)}
\left( n \ell j \; n '\ell' j' \; n ''\ell'' j'' \right)$ (see 
Table~\ref{tab:Implemen_CV1},  \ref{tab:Implemen_CV2}, and \ref{tab:Implemen_CV3}).
These formulae are additionally expressed via the quantities

\begin{equation}
\label{eq:BogAp}
   \mathcal{A^{\prime}}\left(x, \; i j, \; i' j'\right) 
%\nonumber \\ 
  = \sum_{k,k'}
		  \left\{
    \begin{array}{ccc}
      k  & k' & x \\
      j_{i} & j_{i} & j_{i'}
    \end{array} \right\}
				  \left\{
    \begin{array}{ccc}
      k  & k' & x \\
      j_{j} & j_{j} & j_{j'}
    \end{array} \right\}
\mathcal{P}\left(kk', \; i j, \; i' j'\right) ,
\end{equation}
\begin{equation}
\label{eq:BogC}
   \mathcal{C}\left(k, \; i j, \; i' j'\right) 
%\nonumber \\ 
  = \sum_{k'}
		  \left\{
    \begin{array}{ccc}
      k  & j_{i} & j_{i'} \\
      k' & j_{j} & j_{j'}
    \end{array} \right\}
\mathcal{Q}\left(kk', \; i j, \; i' j'\right) ,
\end{equation}
and
\begin{equation}
\label{eq:BogGP}
   \mathcal{G^{\prime}}\left( x_1 \, x_2 \, x, \; i j, \; i' j'\right) 
%\nonumber \\ 
  = \sum_{k,k'}
	   \left( -1 \right)^{k'}
		  \left\{
    \begin{array}{ccc}
      j_{j'} & j_{j'} & x \\
      k      & k'     & j_{j}
    \end{array} \right\}
				  \left\{
    \begin{array}{ccc}
		  j_{i} & j_{i'} & k \\
      x_1    & x_2   & x \\
      j_{i} & j_{i'} & k'
    \end{array} \right\}
\mathcal{P}\left(kk', \; i j, \; i' j'\right) ,
\end{equation}
where
\begin{equation}
\label{eq:BogP}
   \mathcal{P}\left(kk', \; i j, \; i' j'\right) 
   = \mathcal{R}^{k}\left(i j, \; i' j'\right) \; \mathcal{R}^{k'}\left(i' j', \; i j \right)  \; 
	\mathcal{O}\left(K', K \right) ,
\end{equation}

\begin{equation}
\label{eq:BogQ}
   \mathcal{Q}\left(kk', \; i j, \; i' j'\right)
   = \mathcal{R}^{k}\left(i j, \; i' j'\right) \; \mathcal{R}^{k'}\left(i' j', \; j i\right)  \; 
\mathcal{O}\left(K', K \right) ,
\end{equation}

\begin{equation}
\label{eq:BogO1}
\mathcal{O}\left(K', K \right)
= \frac{1}{\overline{E}\left(K' \right)-\overline{E}\left(K\right)} ,
\end{equation}
where $\overline{E}\left(K\right)$ is the averaged energy of the 
state 
for which calculations are performed. $\overline{E}\left(K^{'}\right)$ is the averaged energy for the admixed configuration $K'$.
For details on how to find $\overline{E}\left(K\right)$ and $\overline{E}\left(K^{'}\right)$, see \cite[Section~3]{{Gaigetal:2024CV}}.
We would like to emphasize that the energy denominator (\ref{eq:BogO1}) is defined differently/opposite to the expressions of Feynman diagrams 
(see Fig.~\ref{CV_7}, Eqs. (\ref{eq:BogP}), (\ref{eq:BogQ})).

\begin{eqnarray}
\label{eq:BogC12j}
\hspace*{-0.7cm}
   \mathcal{C}_{12j}\left( i j i', \; k_1 k_2 k_3, \; J_1 J_2 J \right)
\nonumber \\
& & 
\hspace*{-3.2cm}
  = \sum_{x}
	   \left[ x \right]
		  \left\{
    \begin{array}{ccc}
      J_{1} & j_{j} & x \\
      j_{j} & J     & J_{2}
    \end{array} \right\}
		  \left\{
    \begin{array}{ccc}
      J     & j_{j}  & x \\
      k_{3} & j_{i'} & j_{i'}
    \end{array} \right\}
		  \left\{
		 \begin{array}{ccc}
      j_{i'} & k_{3} & x \\
      k_{1}  & j_{i} & k_{2}
    \end{array} \right\}
		  \left\{
		 \begin{array}{ccc}
      j_{i} & k_{1} & x \\
      j_{j} & J_{1} & j_{i}
    \end{array} \right\} .
\end{eqnarray}

We would also like to point out that the notation of ranks such as $J_1$, $J_2$, $y$ in the multipliers of this section under $\Delta \widetilde{\widetilde{\mathcal{R}}}^{(k, k', x)} \left( m n n  \right)$ and $\Delta \widetilde{\widetilde{\mathcal{R}}}^{(k, k', x)} \left( m n p  \right)$ have been renamed for convenience to make the expressions in Table \ref{tab:Implemen_CV2} and \ref{tab:Implemen_CV3} more straightforward and more understandable.

This theory in irreducible tensorial form, as in the papers~\cite{Gaigetal:2024CV,Gaigetal:2024C,Gaigetal:2024CC,Gaigetal:2025VV,Gaigetal:2025VVT}, is more suitable to be included in such version of the {\sc Grasp} which is based on configuration state function generators~\cite{grasp2023,grasp2025}. This is related to the fact that this version of the software package allows us to easily distinguish $F$, $F'$, and $G$ sets of orbitals in the process of computing atomic data. In the following section, we will present a test case of this implementation.

\section{Calculation of core-valence, core, core-core and valence-valence with a new approach}
\label{Sec:Calculations}

In the present work, the method,
based on the Rayleigh-Schr\"odinger perturbation theory in an irreducible tensorial form 
~\cite{Gaigetal:2024CV,Gaigetal:2024C,Gaigetal:2024CC,Gaigetal:2025VV,Gaigetal:2025VVT}, 
is extended to include CV correlations of third and fourth types in the computations. 
We performed MCDHF calculations using the regular method and later using RSMBPT method,  
the CV, C, CC and VV correlations are included in the RCI computations. 
For this purpose, we computed 
the lifetime of the $\mathrm{2s2p^6~^2S_{1/2}}$ state and 
 the 12 lowest energy levels of the $\mathrm{2s^22p^5}$, $\mathrm{2s2p^6}$  
and $\mathrm{2s^22p^4\{3s,3p\}}$ configurations of Ne~II.  
To find the lifetime, only states with $J$ = 1/2 and 3/2 for the odd configurations 
and states with $J$ = 1/2 for the even configurations are needed. 
The multireference (MR) set in the present calculations consists of the 
$\mathrm{2s2p^6}$,  
$\mathrm{2s^22p^4\{3s,3d\}}$, 
$\mathrm{2s2p^4\{3s^2,3d^2\}}$, 
$\mathrm{2s^22p^33p3d}$ and 
$\mathrm{2s^22p^33s3p}$ even as well as  
$\mathrm{2s^22p^5}$, 
$\mathrm{2s^22p^43p}$, 
$\mathrm{2p^43s^23p}$ and 
$\mathrm{2p^43p3d^2}$ odd configurations. 
It should be noted that the MR set in the present computations consists 
of orbitals with few different principal quantum numbers, namely 
with $n$ and $n+1$ (where $n$=2).

\subsection{Computational schemes}
\label{Com_shemes}
The regular MCDHF method was used to calculate the 
radial wave functions of the orbitals in the MR configurations, 
and later the radial functions of the virtual orbitals. 
Only VV correlations were included in this computational scheme
 and it is marked \textbf{VV MCDHF}. 
Single-double (SD) substitutions are allowed from 
$\mathrm{2s,2p_-,2p,3s,3p_-,3p,3d_-,3d}$ valence orbitals of the MR set
to virtual orbital set (OS) $OS_1$=\{$\mathrm{4s,..,4l_{max}}$\}, ..., 
$OS_{5}$=\{$\mathrm{8s,..,8l_{max}}$\}. Three $l_{max}$ values 
were tested: $l_{max} = d,f,g$ in this scheme. 
At the RCI stage, the CV, C and CC correlations were added
 to the investigation, where the $\mathrm{1s}$ orbital is defined as the core. 
This computation scheme was named \textbf{RCI [w VV]}.

The second set of radial functions was computed, including VV and C correlations, 
where the $\mathrm{1s}$ orbital is also defined as the core.  
This computation is marked \textbf{VV+C MCDHF}. Sets of virtual orbitals were 
restricted up to $l_{max} = g$. At the RCI stage, the CV, C and CC correlations were added
 to the investigation as in the previous case, and this scheme was named as \textbf{RCI[w VV+C]}.   

\begin{table}[!h]
{\scriptsize
\setlength{\tabcolsep}{3.5pt}
\caption{Energy levels (in cm$^{-1}$) from NIST ASD ($E_{NIST}$), relative 
differences $(E_{NIST}-E)/E_{NIST}$ (in \%) from NIST ASD 
and our computed energy levels using regular \textbf{RCI[w VV]} - a, \textbf{ RCI[w VV+C]} - b methods, 
and \textbf{RCI (RSMBPT)[w VV]}, \textbf{ RCI (RSMBPT) [w VV+C]}, $rms$ and $rms_3$ (in cm$^{-1}$) 
of all computed levels and of three atomic states involved in the transition that affects lifetimes, 
these states are highlighted. 
}       
\label{Energies_pal}
\centering
\begin{tabular}{l r r r rrrr rrrrr r r}
\hline
\multicolumn{1}{l}{ASF}&\multicolumn{1}{c}{$E_{NIST}$} &\multicolumn{13}{c}{$(E_{NIST}-E)/E_{NIST}$} \\
\cline{3-15}
                                       &        & \multicolumn{1}{c}{a} 
																			                  & \multicolumn{1}{c}{b}  
%-------------------------------------------------------------------------------------------------------------------------	
																			                         &\multicolumn{5}{c}{\textbf{RCI (RSMBPT)[w VV]}} 
																			                                                                      && \multicolumn{5}{c}{\textbf{RCI (RSMBPT)[w VV+C] }}\\ 
																									                  \cline{5-9} \cline{11-15}
																			 &			  &       &       & 95\%  & 99\%  & 99.5\%& 99.95\%&100\% &&  95\%  & 99\%  & 99.5\%& 99.95\% &100\% \\
\hline
\noalign{\smallskip}
\multicolumn{15}{c}{ $l_{max}=d$} \\
\hline
$\mathrm{\bf{2s^22p^5~^2P^o_{3/2} }}$  &      0 &       &       &       &       &       &       &       \\
$\mathrm{2s^22p^4(^3P)3p~^4P^o_{3/2}}$ & 246415 &  0.93 &       &  1.01 &  0.93 &  0.92 &  0.93 &  0.93 \\
$\mathrm{2s^22p^4(^3P)3p~^4D^o_{3/2}}$ & 249696 &  0.80 &       &  0.88 &  0.80 &  0.80 &  0.80 &  0.80 \\
$\mathrm{2s^22p^4(^3P)3p~^2D^o_{3/2}}$ & 251522 &  0.52 &       &  0.57 &  0.50 &  0.50 &  0.51 &  0.52 \\
$\mathrm{\bf{2s^22p^5~^2P^o_{1/2}}}$   &    780 &  1.83 &       & -1.77 & -0.47 &  1.43 &  2.49 &  1.70 \\
$\mathrm{2s^22p^4(^3P)3p~^4P^o_{1/2}}$ & 246598 &  0.94 &       &  1.04 &  0.94 &  0.93 &  0.93 &  0.93 \\
$\mathrm{2s^22p^4(^3P)3p~^4D^o_{1/2}}$ & 249840 &  0.82 &       &  0.89 &  0.82 &  0.82 &  0.82 &  0.82 \\
$\mathrm{2s^22p^4(^3P)3p~^2S^o_{1/2}}$ & 252798 &  0.49 &       &  0.52 &  0.46 &  0.47 &  0.48 &  0.49 \\
$\mathrm{\bf{2s2p^6~^2S_{1/2}      }}$ & 217048 &  1.27 &       &  1.34 &  1.28 &  1.28 &  1.28 &  1.27 \\
$\mathrm{2s^22p^4(^3P)3s~^4P_{1/2}}$   & 219947 &  2.00 &       &  2.18 &  2.04 &  2.02 &  2.00 &  2.01 \\
$\mathrm{2s^22p^4(^3P)3s~^2P_{1/2}}$   & 224699 &  1.89 &       &  2.08 &  1.93 &  1.91 &  1.89 &  1.89 \\
$\mathrm{2s^22p^4(^1S)3s~^2S_{1/2}}$   & 276677 &  1.18 &       &  1.27 &  1.17 &  1.16 &  1.16 &  1.16 \\
\hline                                                              
$rms$                                  &        &  2546 &       &  2771 &  2564 &  2548 &  2538 &  2536 \\
$rms_3$                                &        &  1597 &       &  1673 &  1608 &  1608 &  1603 &  1595 \\
\hline
\multicolumn{15}{c}{$l_{max}=f$} \\                           
\hline                                                
$\mathrm{\bf{2s^22p^5~^2P^o_{3/2} }}$  &      0 &       &       &       &       &       &       &       \\
$\mathrm{2s^22p^4(^3P)3p~^4P^o_{3/2}}$ & 246415 &  0.23 &       &  0.32 &  0.23 &  0.23 &  0.22 &  0.23 \\
$\mathrm{2s^22p^4(^3P)3p~^4D^o_{3/2}}$ & 249696 &  0.17 &       &  0.25 &  0.17 &  0.17 &  0.16 &  0.16 \\
$\mathrm{2s^22p^4(^3P)3p~^2D^o_{3/2}}$ & 251522 & -0.13 &       & -0.07 & -0.15 & -0.15 & -0.14 & -0.14 \\
$\mathrm{\bf{2s^22p^5~^2P^o_{1/2}}}$   &    780 &  1.74 &       & -3.89 & -0.01 & -0.01 &  2.49 &  1.69 \\
$\mathrm{2s^22p^4(^3P)3p~^4P^o_{1/2}}$ & 246598 &  0.24 &       &  0.35 &  0.24 &  0.24 &  0.23 &  0.23 \\
$\mathrm{2s^22p^4(^3P)3p~^4D^o_{1/2}}$ & 249840 &  0.20 &       &  0.26 &  0.19 &  0.19 &  0.19 &  0.19 \\
$\mathrm{2s^22p^4(^3P)3p~^2S^o_{1/2}}$ & 252798 & -0.10 &       & -0.07 & -0.13 & -0.13 & -0.12 & -0.11 \\
$\mathrm{\bf{2s2p^6~^2S_{1/2}      }}$ & 217048 &  0.71 &       &  0.75 &  0.71 &  0.71 &  0.70 &  0.70 \\
$\mathrm{2s^22p^4(^3P)3s~^4P_{1/2}}$   & 219947 &  1.31 &       &  1.50 &  1.36 &  1.36 &  1.31 &  1.30 \\
$\mathrm{2s^22p^4(^3P)3s~^2P_{1/2}}$   & 224699 &  1.26 &       &  1.45 &  1.30 &  1.30 &  1.25 &  1.24 \\
$\mathrm{2s^22p^4(^1S)3s~^2S_{1/2}}$   & 276677 &  0.63 &       &  0.71 &  0.61 &  0.61 &  0.59 &  0.60 \\
\hline                                                             
$rms$                                  &        &  1383 &       &  1590 &  1415 &  1386 &  1366 &  1363 \\
$rms_3$                                &        &   888 &       &   938 &   890 &   883 &   878 &   876 \\
\hline
\multicolumn{15}{c}{ $l_{max}=g$  } \\                          %[w VV]                                 && %[w VV+ C]
\hline                                                          %    95 &   99  & 99.5  & 99.95 & 100   &&    95 &   99  & 99.5  & 99.95 & 100  
$\mathrm{\bf{2s^22p^5~^2P^o_{3/2} }}$  &      0 &       &       &       &       &       &       &       &&       &       &       &       &       \\ 
$\mathrm{2s^22p^4(^3P)3p~^4P^o_{3/2}}$ & 246415 & -0.07 & -0.07 &  0.04 & -0.06 & -0.08 & -0.08 & -0.08 &&  0.03 & -0.06 & -0.08 & -0.08 & -0.08 \\ 
$\mathrm{2s^22p^4(^3P)3p~^4D^o_{3/2}}$ & 249696 & -0.13 & -0.13 & -0.02 & -0.12 & -0.13 & -0.13 & -0.13 && -0.03 & -0.12 & -0.13 & -0.14 & -0.13 \\ 
$\mathrm{2s^22p^4(^3P)3p~^2D^o_{3/2}}$ & 251522 & -0.43 & -0.43 & -0.35 & -0.43 & -0.44 & -0.44 & -0.43 && -0.36 & -0.43 & -0.44 & -0.44 & -0.44 \\ 
$\mathrm{\bf{2s^22p^5~^2P^o_{1/2}}}$   &    780 &  1.65 &  1.61 & -5.86 &  0.54 &  0.61 &  2.79 &  1.60 && -6.56 &  0.68 & -1.24 &  2.63 &  1.57 \\ 
$\mathrm{2s^22p^4(^3P)3p~^4P^o_{1/2}}$ & 246598 & -0.06 & -0.06 &  0.06 & -0.05 & -0.06 & -0.07 & -0.07 &&  0.06 & -0.05 & -0.06 & -0.07 & -0.07 \\ 
$\mathrm{2s^22p^4(^3P)3p~^4D^o_{1/2}}$ & 249840 & -0.10 & -0.10 & -0.02 & -0.09 & -0.10 & -0.11 & -0.11 && -0.03 & -0.09 & -0.10 & -0.11 & -0.11 \\ 
$\mathrm{2s^22p^4(^3P)3p~^2S^o_{1/2}}$ & 252798 & -0.40 & -0.41 & -0.35 & -0.42 & -0.43 & -0.42 & -0.42 && -0.36 & -0.42 & -0.43 & -0.42 & -0.42 \\ 
$\mathrm{\bf{2s2p^6~^2S_{1/2}      }}$ & 217048 &  0.67 &  0.80 &  0.72 &  0.68 &  0.67 &  0.66 &  0.66 &&  0.84 &  0.81 &  0.80 &  0.80 &  0.79 \\ 
$\mathrm{2s^22p^4(^3P)3s~^4P_{1/2}}$   & 219947 &  0.96 &  1.09 &  1.16 &  1.01 &  0.98 &  0.96 &  0.95 &&  1.28 &  1.14 &  1.12 &  1.09 &  1.08 \\ 
$\mathrm{2s^22p^4(^3P)3s~^2P_{1/2}}$   & 224699 &  0.92 &  1.05 &  1.13 &  0.97 &  0.94 &  0.92 &  0.91 &&  1.24 &  1.10 &  1.07 &  1.04 &  1.03 \\ 
$\mathrm{2s^22p^4(^1S)3s~^2S_{1/2}}$   & 276677 &  0.45 &  0.55 &  0.53 &  0.43 &  0.42 &  0.41 &  0.42 &&  0.62 &  0.53 &  0.52 &  0.50 &  0.52 \\ 
\hline  
$rms$                                  &        &  1111 &  1260 &  1265 &  1150 &  1128 &  1104 &  1096 &&  1403 &  1296 &  1276 &  1252 & 1244 \\
$rms_3$                                &        &   836 &  1003 &   901 &   854 &   837 &   831 &   825 &&  1054 &  1019 &  1005 &   999 &  992 \\
\hline                                                                      
\end{tabular}
}
\end{table}

The next investigations were done using the RSMBPT method at the RCI computation. 
The CSF space in this method is divided into three sets: $F$, $F'$ and $G$ 
(see Ref. \cite{Gaigetal:2024CV} for details). 
The $\mathrm{1s}$ subshell is defined as a core subshell 
(that corresponds to the $F$ set), 
$\mathrm{2s,2p_-2p,3s,3p_-,3p,3d_-,3d}$ as valence subshells 
(that correspond to the $F'$ set)  
and subshells belonging to $OS_1$, ..., $OS_{5}$ as virtual ones 
(that correspond to the $G$ set). 
The $l_{max}$ of virtual orbitals 
is described as in the \textbf{VV MCDHF} method. 
This space distribution is consistent with regular {\sc Grasp}2018 calculations
and allows the use of a combination of RCI and RSMBPT methods.

The RSMBPT calculation procedure for RCI is analogous to that used in previous research 
\cite{Gaigetal:2024CV,Gaigetal:2024C,Gaigetal:2024CC,Gaigetal:2025VV,Gaigetal:2025VVT}. 
Using radial functions from \textbf{VV MCDHF} method,   
the contribution of each $K'$ configuration for CSF in the MR set   
was calculated according 
to Rayleigh-Schr\"odinger perturbation theory 
in an irreducible tensorial form, is computed according 
to the 
Eq. (22) of Ref. \cite{Gaigetal:2024CV}   (for first and second types of CV correlations), 
Eq.  (6) of Ref. \cite{Gaigetal:2024C}    (for C correlations),
Eq. (26) of Ref. \cite{Gaigetal:2024CC}   (for CC correlations) 
Eq. (19) of Ref. \cite{Gaigetal:2025VV}   (for first and second types VV correlations),
Eq. (26) of Ref. \cite{Gaigetal:2025VVT}  (for third and fourth VV correlations) and
Eq. (\ref{eq:BogEnergy_PT}) (for third and fourth types of CV correlations). 
$K'$ configurations are sorted in descending order according to the impact of the correlations for each CSF in the MR set. 
Further, $K'$ configurations are selected by CV, C, CC and VV correlations impact with the specified fraction
(expressed in the percentage: 95, 99, 99.5, 99.95 and 100\%) of the total correlations contribution
and then the RCI method was applied, including them.
It should be noted that the program gives the contribution of the correlations of $K'$ configuration 
with a value greater than {\tt 1.0E-11}. Contributions of smaller magnitudes are neglected.
The C correlations (Eq. (3) of Ref. \cite{Gaigetal:2024C}) and V correlations   
which are not included in the RSMBPT method,  
were added to RCI calculations in a regular way.
This computation scheme is named \textbf{RCI (RSMBPT)[w VV]}. 

The RCI (RSMBPT) computation was repeated  
using the radial function from the \textbf{VV+C MCDHF} scheme. 
This scheme was named \textbf{RCI (RSMBPT)[w VV+C]}.   

\begin{table}[h!]
{\scriptsize
\setlength{\tabcolsep}{5pt}
\caption{
Summary of $N_{CSF}$ of standard \textbf{VV MCDHF} at $OS_1$-$OS_5$ and 
\textbf{RCI[w VV]} methods at $OS_5$, \textbf{RCI (RSMBPT)[w VV]} and \textbf{RCI (RSMBPT)[w VV+C]}
method at $OS_5$ is given for $l_{max}=\{d,f,g\}$.}             
\label{NCSF}
\centering
\begin{tabular}{l r rr rrrrrrrrrr}
\hline
\noalign{\smallskip}
\multicolumn{1}{l}{\multirow{2}{*}{$OS$}}       & \multicolumn{2}{c}{$l_{max}=d$} 
															                   && \multicolumn{2}{c}{$l_{max}=f$} 
															 								                   &&\multicolumn{2}{c}{$l_{max}=g$}    &&\multicolumn{2}{c}{$l_{max}=g$}\\
															 \cline{2-3} \cline{5-6} \cline{8-9} \cline{11-12}
                               & odd       & even    &&   odd     & even      &&    odd     & even    &&    odd     & even      \\ 
\hline																																										
\multicolumn{9}{c}{\textbf{VV MCDHF}}                                         &&  \multicolumn{2}{c}{\textbf{VV+C MCDHF}}\\
\cline{2-9} \cline{11-12}
$OS_1$                         &  54~211   &  20~344 &&    89~972 &    35~176 &&    89~972 &    35~176 &&    96~489 &    37~038 \\
$OS_2$                         & 145~282   &  57~840 &&   271~771 &   109~930 &&   345~267 &   137~263 &&   357~960 &   140~955 \\ 
$OS_3$                         & 280~694   & 114~796 &&   552~878 &   226~570 &&   770~998 &   308~520 &&   789~867 &   314~042 \\ 
$OS_4$                         & 460~447   & 191~212 &&   933~293 &   385~096 && 1~367~165 &   548~947 && 1~392~210 &   556~299 \\ 
$OS_5$                         & 684~541   & 287~088 && 1~413~016 &   585~508 && 2~133~768 &   858~544 && 2~164~989 &   867~726 \\ 
\hline
\multicolumn{9}{c}{\textbf{RCI [w VV]}}                                       &&  \multicolumn{2}{c}{\textbf{RCI[w VV+C]}}      \\
\cline{2-9} \cline{11-12}
$OS_5$                         & 2~289~812 & 912~457 && 4~942~682 & 1~877~232 && 7~560~326 & 2~762~366 && 7~560~326 & 2~762~366 \\                                                                                                
\hline
\multicolumn{9}{c}{\textbf{RCI (RSMBPT)[w VV]}}                               &&  \multicolumn{2}{c}{\textbf{RCI (RSMBPT)[w VV+C]}}\\
\cline{2-9} \cline{11-12}    
100\%                          & 2~021~303 & 751~965 && 3~956~501 & 1~343~439 && 5~561~569 & 1~781~337 && 5~561~318 & 1~776~353\\ 
99.95\%                        & 1~854~036 & 689~899 && 3~596~224 & 1~230~114 && 4~885~122 & 1~603~363 && 4~872~057 & 1~602~918\\ 
99.5\%                         & 1~541~262 & 564~815 && 2~790~520 &   954~137 && 3~652~743 & 1~216~955 && 3~636~767 & 1~214~179\\ 
99\%                           & 1~375~125 & 508~995 && 2~431~184 &   821~764 && 3~068~926 & 1~018~785 && 3~045~360 & 1~009~776\\ 
95\%                           &   851~113 & 295~434 && 1~166~070 &   420~313 && 1~360~698 &   484~347 && 1~347~035 &   487~039\\ 
\hline                                                           
\end{tabular}
}
\end{table}

\subsection{Results}
%\subsubsection{Core-valence, core, core-core, and valence-valence correlations}

\subsubsection{Energy levels}
Table \ref{Energies_pal} presents the energy levels recommended by 
the NIST ASD \cite{NIST_ASD} and the relative differences 
$\left( (E_{\mathrm{NIST}} - E) / E_{\mathrm{NIST}} \right)$ between our 
computed energy levels and the NIST ASD-recommended values for the 12 
lowest energy levels of the $\mathrm{2s2p^6}$, $\mathrm{2s^22p^5}$
 and $\mathrm{2s^22p^4\{3s,3p\}}$ configurations. 
Three atomic states involved in the transition that affects lifetimes 
are highlighted.  
The energy levels calculated using the standard 
\textbf{RCI [w VV]} method are presented for the $OS_{5}$ virtual orbital set. 
This set is also restricted by orbital symmetry ($l_{max}={d,f,g}$),  
and the results from each of them are given in the table.
The same table shows the results of further calculations using the 
RSMBPT method. Using this method, the specified amount (95, 99, 99.5, 99.95 and 100\%) of
 CV, C, CC and VV correlations were included in the RCI step. 
These correlations were included up to the virtual orbital sets $OS_{5}$, 
which is as constrained by orbital symmetry 
as in the case of the above MCDHF method.  
Similarly, the relative differences between NIST recommended energy levels and
 energy levels 
obtained using the regular \textbf{RCI [w VV+C]} and \textbf{RCI (RSMBPT) [w VV+C]} 
methods are presented for the $OS_{5}$ virtual orbital set with orbital symmetry $l_{\max}=g$. 
The root-mean-square ($rms$) deviation of all levels 
and those three levels ($rms_3$) involved in transitions depopulating 
$\mathrm{2s2p^6~^2S_{1/2}}$ atomic state are given for all these
 schemes in Table \ref{Energies_pal}.  

The conclusion is that the RCI (RSMBPT) method reproduces 
regular RCI energy levels independently of the orbital symmetry
of virtual orbitals (see line 100\%). In addition, the ability of the RCI (RSMBPT) method
 to reproduce energy levels is independent of the type of correlations included 
in regular MCDHF computations of radial wave functions
 (see line 100\% of \textbf{RCI (RSMBPT) [w VV+C]} column).
It should be noted that, for this ion, the RCI (RSMBPT) 
method already reproduces the regular RCI results for most
 energy levels when 99.5\% of the correlations are included; 
only the $\mathrm{2s^22p^5~^2P^o_{1/2}}$ level requires full (100\%) correlation inclusion.

\begin{figure}
\centering
\includegraphics[width=0.45\textwidth]{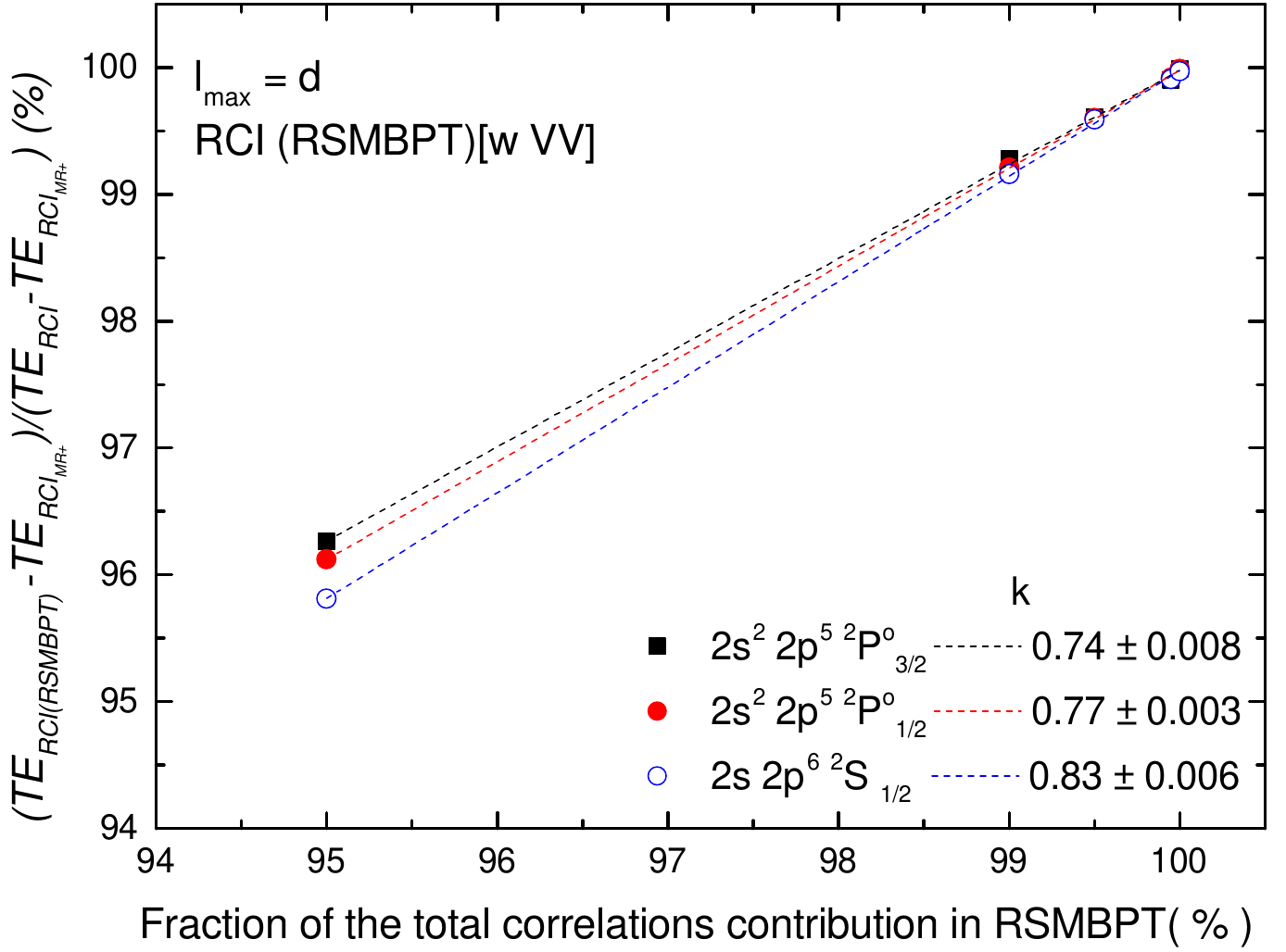}
\includegraphics[width=0.45\textwidth]{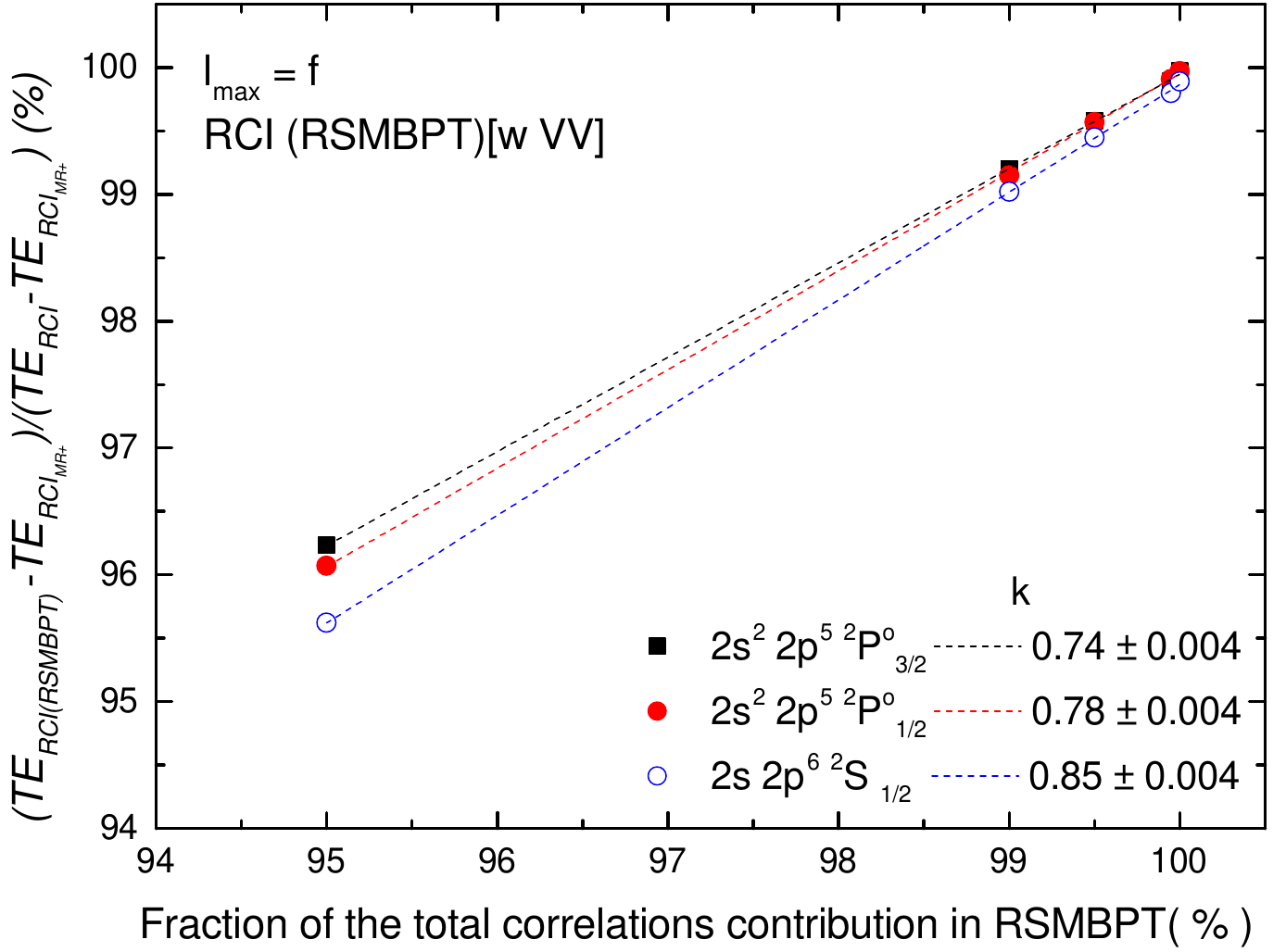} \\
\includegraphics[width=0.45\textwidth]{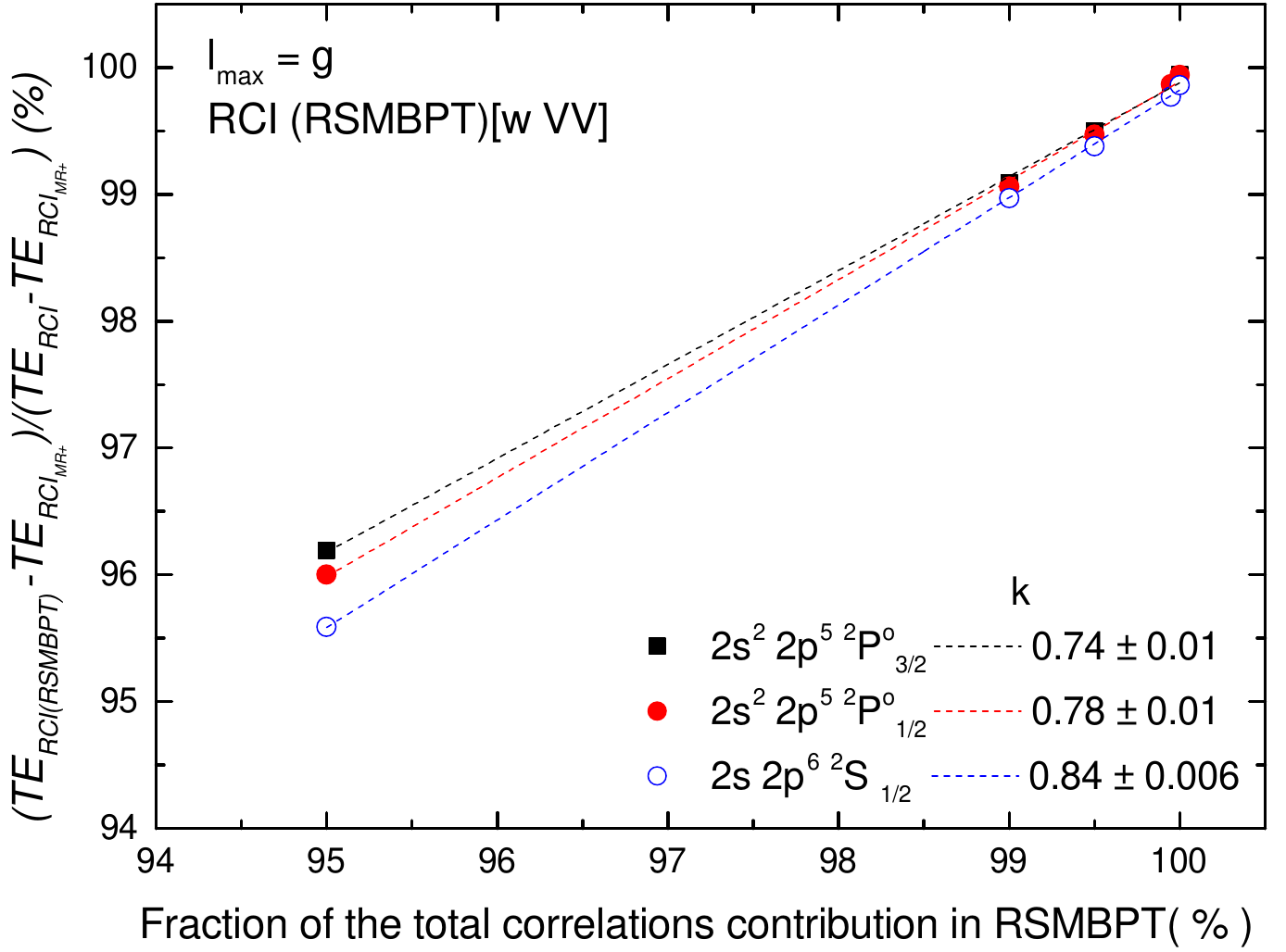}
\includegraphics[width=0.45\textwidth]{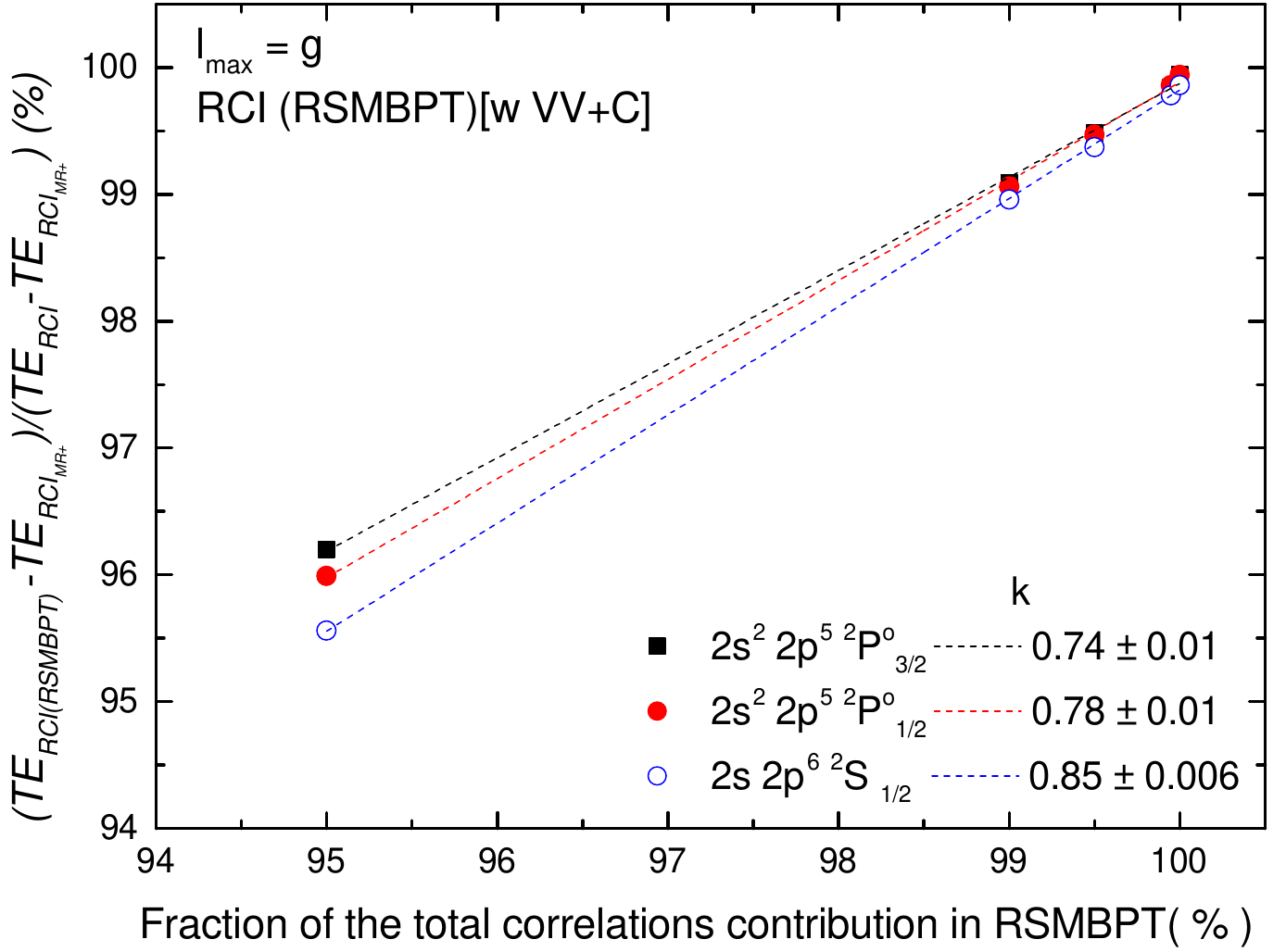}
\caption{\label{convergence_PT} 
Dependence of the included correlation contribution (CV+C+CC+VV)
 on the specified fraction of the total correlation contribution 
used in the RCI (RSMBPT) method for the three lowest levels,
 for $l_{\max}=d,f,g$, and for radial wave functions computed with VV+C correlations.}  
\end{figure}

Figure \ref{convergence_PT} illustrates the contribution of the included correlations 
($\Delta_{\mathrm{cor}}$) obtained using the RCI (RSMBPT) method, 
relative to the complete standard RCI calculations, 
as a function of the fraction of the total correlation contribution included in the RCI (RSMBPT) approach.
This dependence is shown for the $\mathrm{2s^22p^5~^2P^o_{3/2}}$, $\mathrm{2s^22p^5~^2P^o_{1/2}}$ and 
$\mathrm{2s2p^6~^2S_{1/2}}$ levels when CV, C, CC and VV correlations are taken into account. 
The same dependence is presented for each tested orbital symmetry
 of the virtual space, $l_{\max}=d,f,g$, and for radial wave functions
 computed with VV+C correlations included.   
The contribution $\Delta_{\mathrm{cor}}$ (in \%) is defined as
$(TE_{RCI (RSMBPT)} - TE_{\textbf{MR+}}) / (TE_{RCI} - TE_{\textbf{MR+}})$,
where $TE$ denotes the total energy obtained within a given computational scheme.
Here, $TE_{RCI (RSMBPT)}$ is the total energy calculated using the RCI (RSMBPT) 
method with a specified fraction (in \%) of the total correlation contribution included.
The $TE_{\textbf{MR+}}$ values are obtained from calculations in which the CSF basis
 consists of the MR set augmented by correlation effects not included in the RSMBPT method. 
As can be observed in the figure, the computed $\Delta_{cor}$ 
is almost identical to the specified fraction (in \%) 
of the total correlations contribution used in the RCI (RSMBPT) method, exhibiting a linear dependence, 
independent of the orbital symmetry and the correlations included in the radial function calculations. 
The linear fitting ($y = kx + a$) demonstrates that the slope coefficient ($k$) is less than unity, ranging from 0.74 to 0.85, 
and that the error associated with the fitting itself is negligible compared to the resulting values (see Fig. \ref{convergence_PT}), 
indicating that the $k$-factor is highly precise.
Based on these observations, it can be concluded 
that the difference between the correlation value 
(in percent) in the predefined RSMBPT calculations
 and 100\% indicates the percentage error in the total energy, 
which is defined as the difference between the 
total energy calculated using the regular RCI method 
(or using the 100\% correlations in RSMBPT method) 
and the total energy calculated using the reduced correlation
 obtained due to the shortening of the CSF base. 
For example, using the RSMBPT method total energies can be analyzed 
 from reduced CSF bases 
(e.g., 95\% and 99\% correlation inclusion), it is possible to 
accurately extrapolate the total energy that would be obtained 
when 100\% of the correlations are included.
The extrapolation of the total energies
would be helpful for complex computations when
considering CV, C, CC and VV correlations, as
such computations lead to a large CSFs basis and
are time consuming. 
Moreover, these regularities remain stable across 
different orbital symmetries of the virtual space and
 for different types of correlations included in 
the radial wave function calculations.

Moreover, these regularities remain stable across different orbital symmetries of the virtual space and
 for different types of correlations included in the radial wave function calculations.
It is worth noting that this theory has been developed 
only recently; therefore, it is important to investigate this dependence in greater detail.   

Table \ref{NCSF} gives a summary of CSF numbers ($N_{CSFs}$) for 
a regular \textbf{VV MCDHF} and \textbf{VV+C MCDHF} 
computations at $OS_{1-5}$, for
 all tested orbitals' symmetries $l_{max}={d,f,g}$. 
Also, CSF numbers are given for 
\textbf{RCI[w VV]}, \textbf{RCI[w VV+C]}, 
\textbf{RCI (RSMBPT)[w VV]} and \textbf{RCI (RSMBPT)[w VV+C]} at $OS_5$ for
 all tested orbitals' symmetries. 
For the RCI (RSMBPT) method, the number of CSF is given, based on 
the amount: 95, 99, 99.5, 99.95 and 100\% of CV, C, CC and VV correlations.
  
As can be seen from the table, even counting 
 the simplest type of correlations 
i.e. VV electron correlations with increasing $OS$, $N_{CSFs}$ grows very fast, 
even without including very high orbital symmetries, such as $h$ or higher  
(see $N_{CSFs}$ at \textbf{VV MCDHF} at $OS_5$ with orbital symmetries up to $g$). 
The inclusion of CV, C and CC correlations of the $\mathrm{1s}$ orbital increases the
 size of the bases by a factor of 3.2-4.3 
(see data of the \textbf{RCI [w VV]} and \textbf{RCI [w VV+C]} methods). 
Meanwhile, the RCI (RSMBPT) method 
reduces $N_{CSFs}$ by 12\%-36\%, 
even at the case where 100\% of correlations are included 
and this reduction is more prominent in larger $l$ symmetries. 
This reduction is crucial for performing highly accurate calculations, 
which often require high orbital symmetries and consequently lead to very large CSF bases.

\begin{table}[h!]
{\scriptsize
\setlength{\tabcolsep}{1pt}
\caption{
Line strengths (in a.u.) in Babushkin gauge ($S_B$) for two transitions: 
$\mathrm{2s2p^6~^2S_{1/2}}-\mathrm{2s^22p^5~^2P^o_{3/2}}$ (3-1) and   
$\mathrm{2s2p^6~^2S_{1/2}}-\mathrm{2s^22p^5~^2P^o_{1/2}}$ (3-2) %(see text for details) 
computed using standard \textbf{VV MCDHF}, \textbf{VV+C MCDHF}, 
\textbf{RCI [w VV]} and \textbf{RCI[w VV+C]} methods, and \textbf{RCI (RSMBPT)[w VV]} and \textbf{RCI (RSMBPT)[w VV+C]} methods. }             
\label{ACC_for_S}
\centering
\begin{tabular}{llllllllllll }
\hline\hline
\multicolumn{1}{l}{$OS$}       & \multicolumn{11}{c}{$S_B$}\\
\cline{2-12}               
                               & \multicolumn{2}{c}{$l_{max}=d$} 
															                               && \multicolumn{2}{c}{ $l_{max}=f$} 
																								                   &&\multicolumn{2}{c}{$l_{max}=g$} &&\multicolumn{2}{c}{$l_{max}=g$}\\
\cline{2-3}\cline{5-6}\cline{8-9}\cline{11-12}				
                               & 3-1          & 3-2          && 3-1          & 3-2          && 3-1          & 3-2     && 3-1          & 3-2 \\
\hline
\noalign{\smallskip}
                               & \multicolumn{8}{c}{\textbf{VV MCDHF}} && \multicolumn{2}{c}{\textbf{VV+C MCDHF}} \\
\cline{2-9}\cline{11-12}
$OS_1$                         & 3.439E-01 AA & 1.537E-01 AA && 5.156E-01 A+ & 2.393E-01 A+ && 5.156E-01 A+ & 2.393E-01 A+ && 6.192E-01 A  & 2.931E-01 A  \\
$OS_2$                         & 5.412E-01 A  & 2.658E-01 A  && 5.471E-01 A+ & 2.688E-01 A  && 5.568E-01 A+ & 2.729E-01 A  && 5.356E-01 A  & 2.647E-01 A  \\ 
$OS_3$                         & 5.406E-01 A  & 2.648E-01 A  && 5.419E-01 A+ & 2.649E-01 A  && 5.489E-01 A+ & 2.680E-01 A  && 5.411E-01 A  & 2.651E-01 A  \\ 
$OS_4$                         & 5.387E-01 A  & 2.623E-01 A  && 5.308E-01 A+ & 2.578E-01 A  && 5.362E-01 A+ & 2.605E-01 A  && 5.317E-01 A  & 2.584E-01 A  \\ 
$OS_5$                         & 5.385E-01 A  & 2.623E-01 A  && 5.307E-01 A+ & 2.579E-01 A  && 5.363E-01 A+ & 2.606E-01 A+ && 5.315E-01 A  & 2.584E-01 A  \\ 
\hline                                                             % l_max_g_bePT
                               &\multicolumn{8}{c}{\textbf{RCI [w VV]}}          && \multicolumn{2}{c}{\textbf{RCI[w VV+C]}}\\
\cline{2-9}\cline{11-12}
$OS_5$                         & 5.496E-01 B+ & 2.680E-01 B+ && 5.384E-01 AA & 2.625E-01 AA && 5.414E-01 AA & 2.645E-01 AA && 5.415E-01 AA & 2.645E-01 AA \\
\hline
                               &\multicolumn{8}{c}{\textbf{RCI (RSMBPT)[w VV]}}                                            
															                                                                                             && \multicolumn{2}{c}{\textbf{RCI (RSMBPT)[w VV+C]}} \\
\cline{2-9}\cline{11-12}       %&        M    &           M  &&       M      & M            &&           M  &        M     &&         M    &       M        
100\%                          & 5.498E-01 B+ & 2.681E-01 B+ && 5.387E-01 AA & 2.627E-01 AA && 5.418E-01 AA & 2.647E-01 AA && 5.419E-01 AA & 2.647E-01 AA \\
99.95\%                        & 5.503E-01 B+ & 2.681E-01 B+ && 5.393E-01 AA & 2.627E-01 AA && 5.424E-01 AA & 2.647E-01 AA && 5.425E-01 AA & 2.647E-01 AA \\
99.5\%                         & 5.506E-01 B+ & 2.684E-01 B+ && 5.397E-01 AA & 2.629E-01 AA && 5.429E-01 AA & 2.648E-01 AA && 5.431E-01 AA & 2.648E-01 AA \\
99\%                           & 5.507E-01 B+ & 2.684E-01 B+ && 5.404E-01 A+ & 2.633E-01 AA && 5.433E-01 AA & 2.652E-01 AA && 5.434E-01 AA & 2.653E-01 AA \\
95\%                           & 5.495E-01 B+ & 2.684E-01 B+ && 5.399E-01 A  & 2.631E-01 A+ && 5.436E-01 A  & 2.654E-01 A+ && 5.440E-01 A  & 2.654E-01 AA \\
\hline
\end{tabular}
}
\end{table}

%\subsubsection{\textcolor[rgb]{0,0,1}{Quantitative and qualitative evaluation}}
\subsubsection{Quantitative and qualitative evaluation}
\begin{figure*}
\centering
\includegraphics[width=0.45\textwidth]{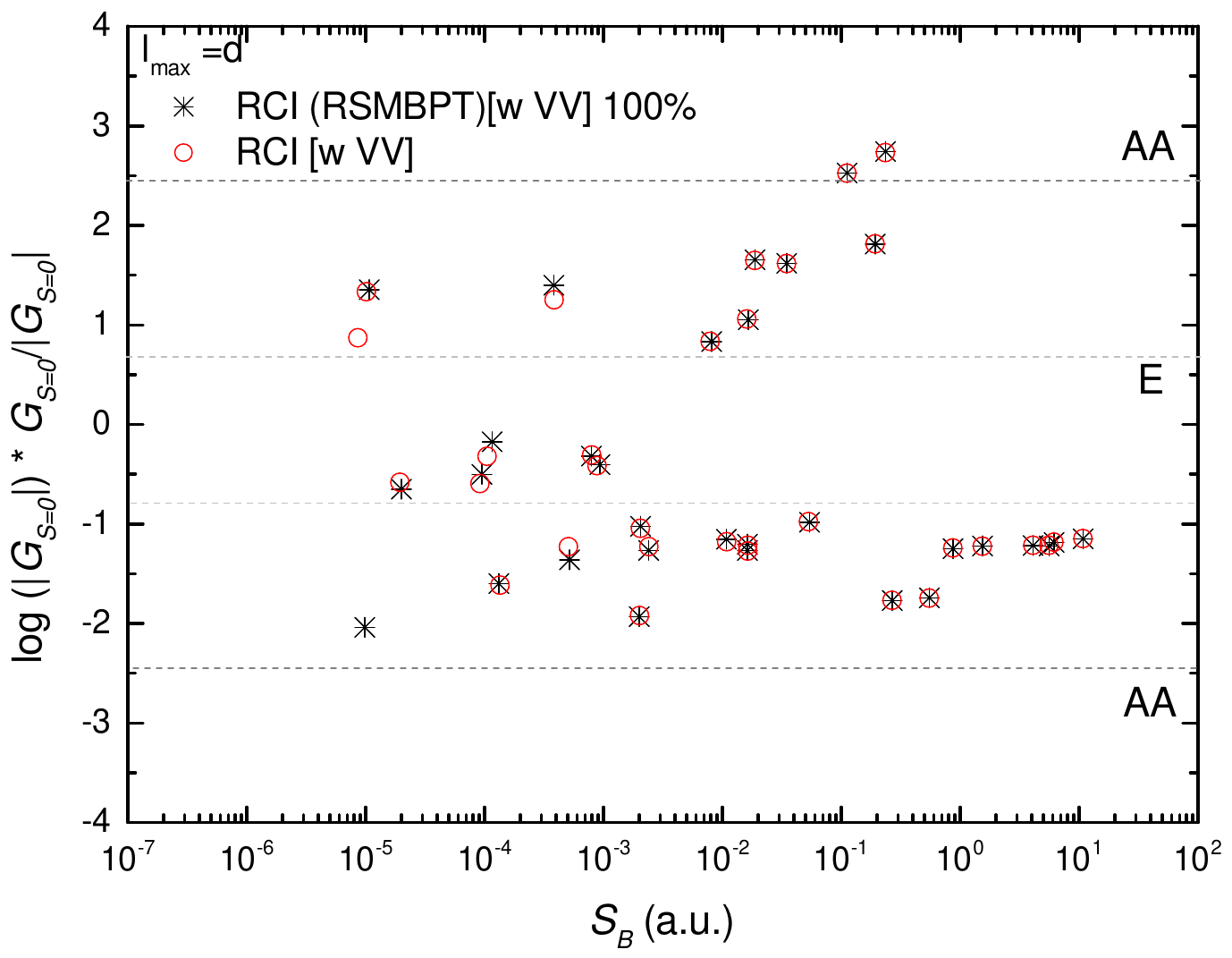}
\includegraphics[width=0.45\textwidth]{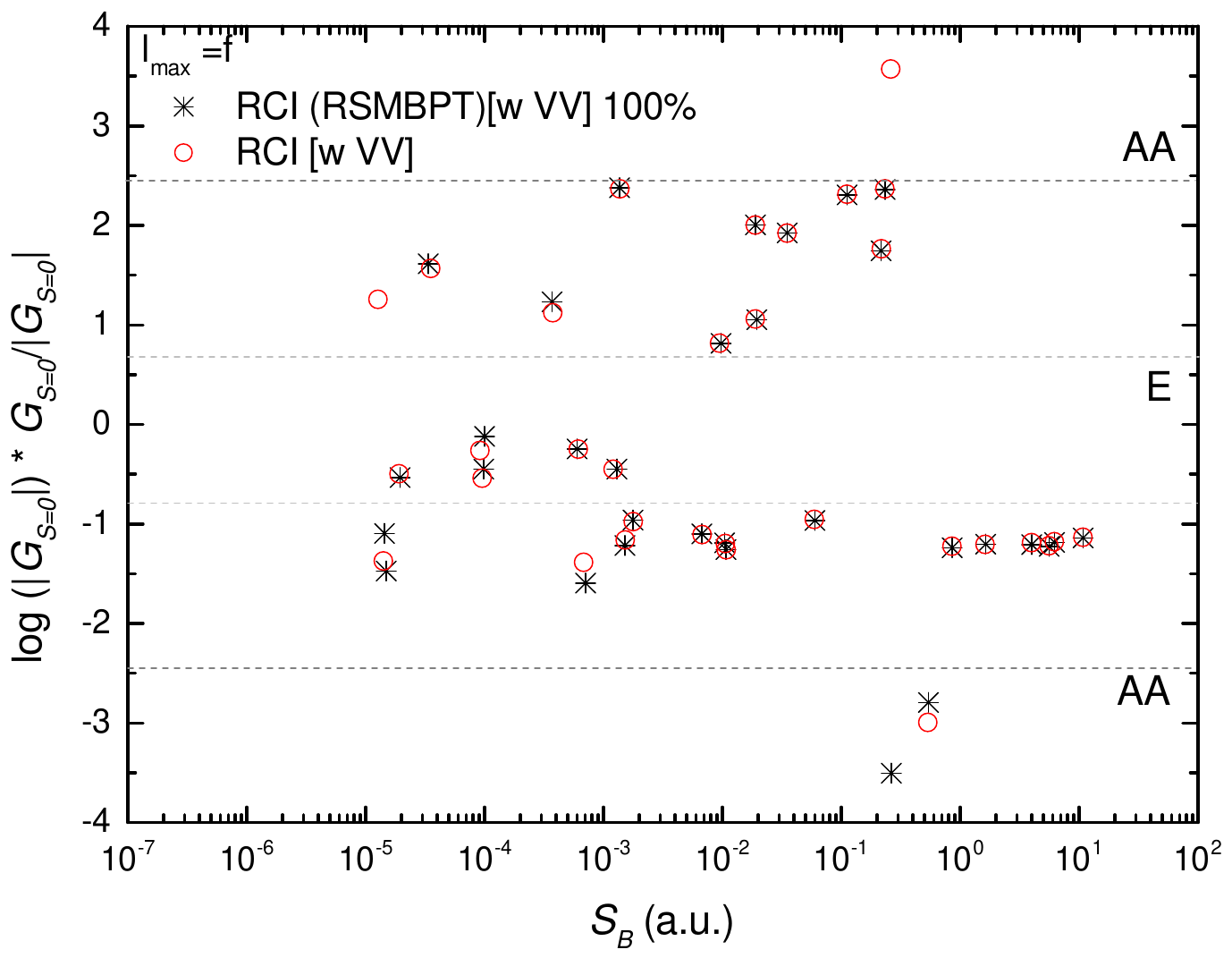} \\
\includegraphics[width=0.45\textwidth]{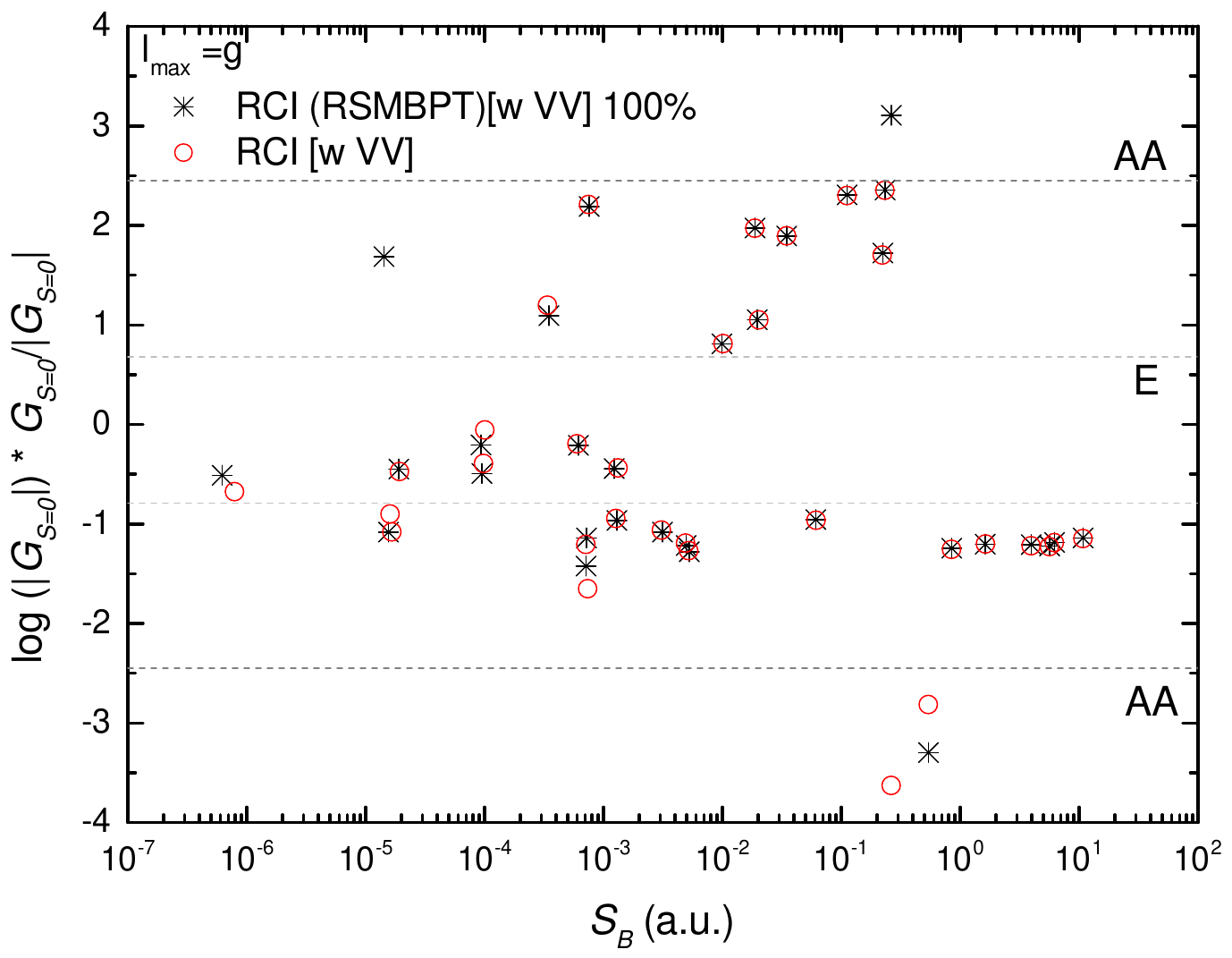}
\includegraphics[width=0.45\textwidth]{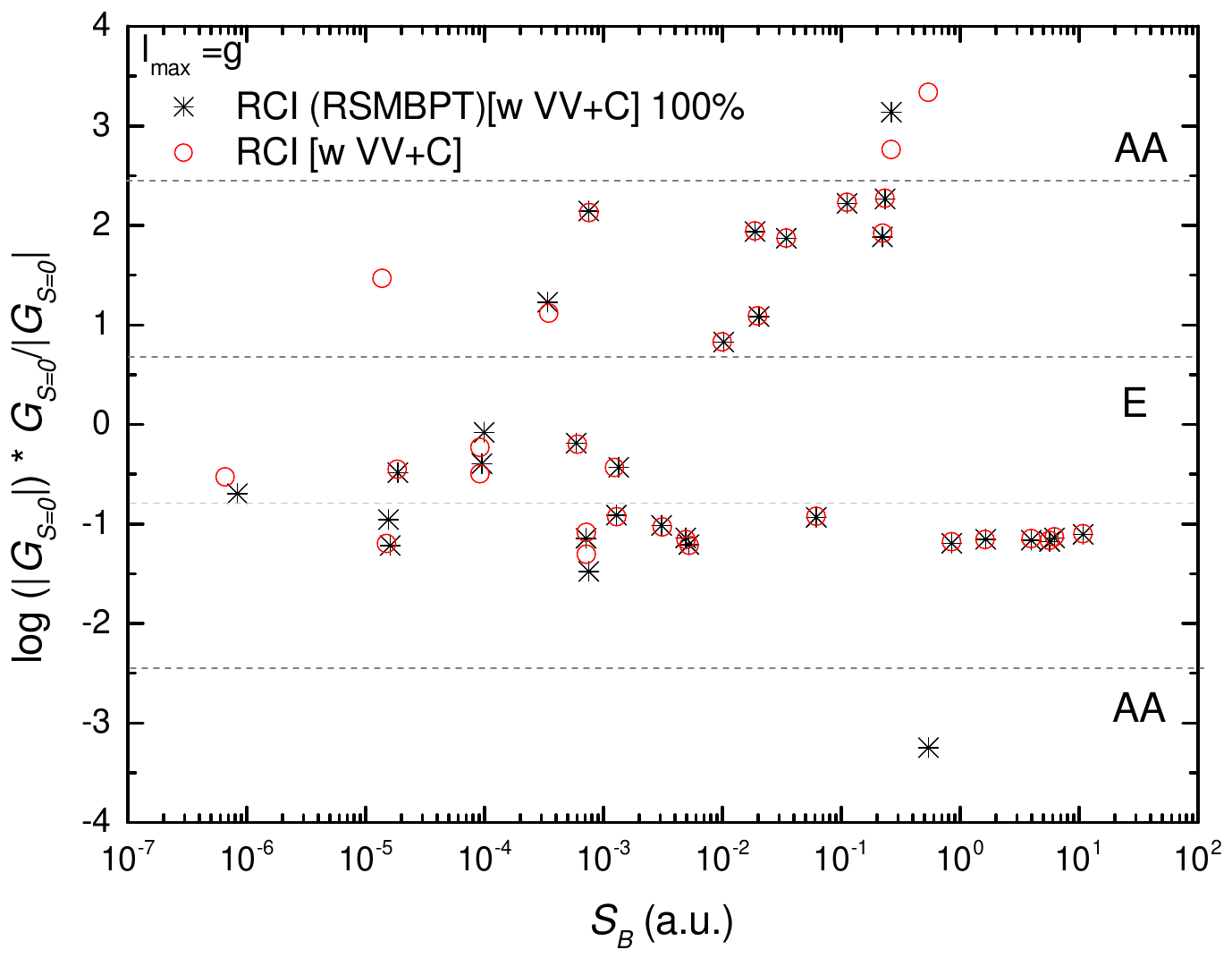}
\caption{\label{S_x_min} 
The $G_{S=0}$ values for each line strength computed using the standard \textbf{RCI [w VV]} method
 and the \textbf{RCI (RSMBPT) [w VV]} method with 100\% of the correlations included for $l_{\max}={d,f,g}$ and 
 the \textbf{RCI [w VV+C]}
 and \textbf{RCI (RSMBPT) [w VV+C]} schemes with 100\% of the correlations included.}  
\end{figure*}

The transition properties of 32 E1-type transitions between 
levels of even configurations with $J = 1/2$ and levels of odd configurations with $J = 1/2$ and $3/2$ were computed.
The calculations are done in a regular way and using 
the RSMBPT method when CV, C, CC and VV are included. 
The quantitative and qualitative evaluation method (QQE)~\cite{Se_Ge_like,Ce_IV,Pr_IV} 
has been used to investigate the uncertainty of the line strengths ($S$). 
Within the QQE method, the minimum of the parabola $G_{S=0}$ \cite{Se_Ge_like} values were determined for 
each transition and subsequently used to assign the corresponding accuracy classes.
The accuracy classes match   
the NIST ASD~\citep{NIST_ASD} terminology (AA $\leq$ 1~\%, A${+}$ $\leq$ 2~\%, A $\leq$ 3~\%, 
B${+}$ $\leq$ 7~\%, B $\leq$ 10~\%, C${+}$  $\leq$ 18~\%,  C $\leq$ 25~\%,  
D${+}$ $\leq$ 40~\%, D $\leq$ 50~\%, and E $>$ 50~\%).
The $G_{S=0}$ values for each line strength computed using the standard \textbf{RCI [w VV]} method
 and the \textbf{RCI (RSMBPT) [w VV]} method with 100\% of the correlations included 
are shown in Fig. \ref{S_x_min} for $l_{\max}={d,f,g}$. 
The figure also includes results obtained using radial wave functions 
that incorporate both C and VV correlations, namely the \textbf{RCI [w VV+C]}
 and \textbf{RCI (RSMBPT) [w VV+C]} schemes with 100\% of the correlations included.  
These correlations were included up to the virtual orbital sets $OS_{5}$.

As can be seen from Fig. \ref{S_x_min}, the most $S$ values calculated using the RCI (RSMBPT)
 method with 100\% of the correlations included coincide with 
those obtained using the standard RCI method in most cases, 
independently of the orbital symmetry and of the correlations included
 in the calculation of the radial wave functions. Noticeable differences 
are observed only for the smallest $S$ values.

Figure \ref{S_x_min} also shows that when using the 
RCI (RSMBPT) method with 100\% correlations, 
in most cases $G_{S=0}$ values are obtained, 
which correspond to those obtained using the standard RCI method.
Small differences are observed for 
the smallest $S$ values, while larger deviations appear for 
the largest $G_{S=0}$ values, which correspond to the 
highest accuracy classes. In these cases, the sign of the 
$G_{S=0}$ value changes; however, the accuracy class remains unchanged. 

Table \ref{ACC_for_S} presents the line strengths (in the Babushkin gauge) for two transitions,
$\mathrm{2s2p^6~^2S_{1/2}} \rightarrow \mathrm{2s^22p^5~^2P^o_{3/2}}$ (3-1) and
$\mathrm{2s2p^6~^2S_{1/2}} \rightarrow \mathrm{2s^22p^5~^2P^o_{1/2}}$ (3-2),
which contribute to the lifetime under consideration. 
The line strengths were calculated using the standard 
\textbf{RCI [w VV]} and the \textbf{RCI (RSMBPT) [w VV]} 
methods including virtual orbitals up to $OS_5$ for $l_{\max}={d,f,g}$.
Additionally, results obtained with radial functions that 
include C and VV correlations are reported for the 
\textbf{RCI [w VV+C]} and \textbf{RCI (RSMBPT) [w VV+C]} schemes.
The corresponding accuracy classes are also provided.
The line strengths calculated using the RCI method coincide
 with those obtained using the RCI (RSMBPT) method when 
100\% correlation is included, with differences at the 0.0-0.1\% level, 
independently of the orbital symmetry and the 
correlations included in the radial function calculations.
As the amount of included correlation decreases, the agreement
 with the standard RCI results slightly deteriorates; 
however, the relative differences,
$\left(S_{\mathrm{RCI}} - S_{\mathrm{RCI(RSMBPT)_{95\%}}}\right)/S_{\mathrm{RCI}}$,
remain below 0.5\%. 
The accuracy classes of these line strengths, 
computed using the RCI method, 
coincide with those obtained using the RCI (RSMBPT)
approach when 100\% correlation is included, 
independently of the orbital symmetry and the 
correlations included in the radial function calculations.
As the percentage of included correlation decreases,
 the accuracy slowly deteriorates.
 
\begin{table}[h!]
{\scriptsize
\setlength{\tabcolsep}{9pt}
\caption{
Lifetime (in ns) of the $\mathrm{2s2p^6~^2S_{1/2}}$ state is computed using standard \textbf{VV MCDHF}, \textbf{VV+C MCDHF}, 
\textbf{RCI [w VV]}, \textbf{RCI[w VV+C]} method, and \textbf{RCI (RSMBPT)[w VV]}, \textbf{RCI (RSMBPT)[w VV+C]} methods. 
Lifetimes are given in the Babushkin and in the Coulomb (in brackets) gauges.}             
\label{lifetimes_2}
\centering
\begin{tabular}{l r r r rrr }
\hline\hline
\multicolumn{1}{l}{$OS$}       & \multicolumn{3}{c}{$\tau$}\\
\cline{2-7}
                               & \multicolumn{1}{c}{$l_{max}=d$} 
															                   & \multicolumn{1}{c}{ $l_{max}=f$} 
																								                   &\multicolumn{1}{c}{$l_{max}=g$} &&\multicolumn{1}{c}{$l_{max}=g$}\\
																																	
\hline
\noalign{\smallskip}
\multicolumn{4}{c}{\textbf{VV MCDHF}} && \multicolumn{2}{c}{\textbf{VV+C MCDHF}} \\
\cline{2-4}\cline{6-7}                                                          
MR                             & 0.1171 (0.09761)& 0.1171 (0.09761)& 0.1171 (0.09761)&& 0.1171 (0.09761)\\
$OS_1$                         & 0.1991 (0.2007) & 0.1991 (0.2007) & 0.1287 (0.1262) && 0.1066 (0.1042) \\
$OS_2$                         & 0.1223 (0.1258) & 0.1187 (0.1165) & 0.1165 (0.1140) && 0.1207 (0.1180) \\ 
$OS_3$                         & 0.1223 (0.1259) & 0.1198 (0.1175) & 0.1182 (0.1156) && 0.1196 (0.1169) \\ 
$OS_4$                         & 0.1229 (0.1265) & 0.1227 (0.1202) & 0.1213 (0.1186) && 0.1222 (0.1195) \\ 
$OS_5$                         & 0.1229 (0.1264) & 0.1227 (0.1202) & 0.1212 (0.1185) && 0.1222 (0.1195) \\  
\hline                                                             
                                            \multicolumn{4}{c}{\textbf{RCI [w VV]}}  && \multicolumn{2}{c}{\textbf{RCI[w VV+C]}}\\
\cline{2-4}\cline{6-7} 
$OS_5$                         & 0.1231 (0.1293) & 0.1236 (0.1238) & 0.1226 (0.1227) && 0.1231  (0.1228) \\
\hline
                                            \multicolumn{4}{c}{\textbf{RCI (RSMBPT)[w VV]}}
																						                                         && \multicolumn{2}{c}{\textbf{RCI (RSMBPT)[w VV+C]}} \\
\cline{2-4}\cline{6-7}                                                                                        
100\%                          & 0.1231 (0.1293) & 0.1235 (0.1239) & 0.1225 (0.1229) && 0.1230 (0.1231) \\ 
99.95\%                        & 0.1230 (0.1297) & 0.1234 (0.1242) & 0.1224 (0.1233) && 0.1229 (0.1234) \\ 
99.5\%                         & 0.1230 (0.1300) & 0.1233 (0.1244) & 0.1224 (0.1234) && 0.1228 (0.1235) \\ 
99\%                           & 0.1230 (0.1302) & 0.1232 (0.1244) & 0.1223 (0.1234) && 0.1228 (0.1235) \\ 
95\%                           & 0.1233 (0.1321) & 0.1235 (0.1239) & 0.1224 (0.1249) && 0.1228 (0.1250) \\    
\hline                                                           
Ex. \cite{Koike}               & 0.132 \\
Th. \cite{Koike}               & 0.120 (0.126)\\ 
Th. \cite{Charlotte}           & 0.1214\\ 
\hline
\end{tabular}
}
\end{table}

\subsubsection{Lifetime}
Table \ref{lifetimes_2} presents the lifetime of the $\mathrm{2s2p^6~^2S_{1/2}}$ state.
The lifetime was calculated using the standard MCDHF method,
 which includes only VV correlations in different virtual orbital sets ($OS_1$ through $OS_5$), 
restricted by orbital symmetry ($l_{\max} = {d,f,g}$).
The table also presents results obtained using the RCI and RCI (RSMBPT) methods 
 with the virtual orbital setup to $OS_5$, under the 
same orbital-symmetry constraints as in the MCDHF calculations. 
Various percentages of CV, C, CC and VV correlations (95, 99, 99.5, 99.95 and 100\%) 
were included in the RCI (RSMBPT) calculations.
Lifetimes are reported in the Babushkin gauge, with values in the Coulomb gauge given in brackets.

The lifetime in Babushkin gauge of the $\mathrm{2s2p^6~^2S_{1/2}}$ state is reproduced, using the RCI (RSMBPT)
method in the case of the ”100\%”, with a precision of 4 digits 
(i.e., $\tau_{RCI(RSMBPT)}-\tau_{RCI} = 0.0001$)
compared to the regular {\sc Grasp}2018 calculations, 
 irrespective of the orbital
 symmetries that are included in the calculations and 
of the types of correlations included in radial function computations.  
 The lifetime in Coulomb’s gauge is reproduced with a precision of 4 digits 
(i.e., $\tau_{RCI(RSMBPT)}-\tau_{RCI} \leq 0.0003$). 

The lifetime is also very stable with decreasing correlation percentage 
 regardless of the orbital symmetry included.  
The difference in lifetimes between the 100\% and 95\% cases is also small 
($\tau_{RCI(RSMBPT_{100\%})}-\tau_{RCI(RSMBPT_{95\%})} \leq 0.0002$ 
Babushkin's gauge and $\leq 0.0028$ for Coulomb’s gauge). 

The experimental \cite{Koike} 
and theoretical \cite{Koike,Charlotte} lifetimes are given for comparison.  
Experiment was done using the time-correlated single
photon counting technique combined with photoionization by synchrotron radiation. 
Theoretical computations for lifetime are done 
using the multi-configuration Dirac-Fock method in the same work \cite{Koike}. 
The lifetimes from the last investigation are given in the Babushkin and in the Coulomb (in brackets) 
gauges. The next theoretical work \cite{Charlotte} was done using the multi-configuration
Hartree-Fock method with relativistic effects 
included through the Breit-Pauli Hamiltonian, omitting only the orbit-orbit
interaction. 

Our calculated lifetime using the 
RCI (RSMBPT) and regular RCI methods 
is smaller than the experimentally measured value \cite{Koike}, 
but slightly bigger than the theoretical evaluation 
by \cite{Koike} and \cite{Charlotte}.

\section{Conclusions}
\label{Sec:Conclusions}
The method, based on the Rayleigh-Schr\"odinger perturbation 
theory in an irreducible tensorial form, 
is extended to estimate the CV correlations of the third and fourth types.
We provided the analytical expression of a three-particle Feynman diagram describing these CV
correlations with explanations and expression for the spin-angular coefficients.
The expressions to calculate the influence of these correlations are derived and provided.
This extended RSMBPT method allows us to 
evaluate the contribution of each $K'$ configuration, 
which determines one of the CV, C, CC, or VV correlation types, 
to the CSFs in the MR set. 
The method applies to the investigation of correlations
in arbitrary electronic configurations of atoms or ions, 
using selected core and virtual orbital sets and allowing for any number of valence electrons.
The 12 lowest energy levels of the $\mathrm{2s^22p^5}$, $\mathrm{2s2p^6}$   
and $\mathrm{2s^22p^4\{3s,3p\}}$ configurations, transition properties of 32 E1-type between these levels, and  
the lifetime of $\mathrm{2s2p^6~^2S_{1/2}}$ state
were computed for Ne~II.   

The analysis of the energy levels has shown 
that the RCI (RSMBPT) method reproduces the 
regular RCI energy levels independently of 
the orbital symmetry of the virtual orbitals. 
In addition, the ability of the RCI (RSMBPT) 
method to reproduce the energy levels is independent 
of the type of correlations included in the regular MCDHF computations.

Using the RSMBPT method total energies can be analyzed 
 from reduced CSF bases 
(e.g., 95\% and 99\% correlation inclusion), it is possible to 
accurately extrapolate the total energy that would be obtained 
when 100\% of the correlations are included.
The extrapolation of the total energies
would be helpful for complex computations when
considering CV, C, CC and VV correlations, as
such computations lead to a large CSFs basis and
are time consuming. 
Moreover, these regularities remain stable across 
different orbital symmetries of the virtual space and
 for different types of correlations included in 
the radial wave function calculations.

The accuracy classes of the line strengths for 
transitions originating from the $\mathrm{2s2p^6~^2S_{1/2}}$ state 
coincide, when 100\% of correlations were included in  
the RCI (RSMBPT) [w VV] and \textbf{RCI (RSMBPT) [w VV+C]} schemes. 
As the percentage of included correlations decreases, 
the accuracy classes gradually deteriorate.

The lifetime of the $\mathrm{2s2p^6~^2S_{1/2}}$ 
state is reproduced using the RCI (RSMBPT) 
method with 100\% of the correlations included, 
with a precision of four digits 
(i.e., $\tau_{\mathrm{RCI(RSMBPT)}} - \tau_{\mathrm{RCI}} = 0.0001$ 
in the Babushkin gauge and $\leq 0.0003$ in the Coulomb gauge), 
compared to the regular {\sc Grasp}2018 calculations, 
independently of the orbital symmetries included 
in the calculations and of the types of correlations 
included in the radial wave-function computations.
The lifetime is also very stable with decreasing correlation percentage 
 regardless of the orbital symmetry inclusion. 
The difference in lifetimes in the cases 100\% and 95\% is also small 
($\tau_{RCI(RSMBPT_{100\%})}-\tau_{RCI(RSMBPT_{95\%})} \leq 0.0002$ 
Babushkin's gauge and $\leq 0.0028$ for Coulomb’s gauge). 
The same stability of the lifetime value is observed 
when the radial functions are calculated 
by including the VV+C correlation.

\end{document}